\documentclass[final,1p,times]{elsarticle}
\usepackage{amssymb}
\usepackage{amsfonts}
\usepackage{amsmath}
\usepackage{geometry}
\usepackage{bm}
\numberwithin{equation}{section}

\newdefinition{rmk}{Remark}
\newproof{pf}{Proof}
\newproof{pot}{Proof of Theorem \ref{thm2}}

\begin{document}

\begin{frontmatter}
\title{Approximative solution of the spin free Hamiltonian involving only scalar potential for the $q-\bar{q}$ system}

\author{Yoon Seok Choun\corref{cor1}}
\ead{Yoon.Choun@baruch.cuny.edu; ychoun@gmail.com}
\cortext[cor1]{Correspondence to: Baruch College, The City University of New York, Natural Science Department, A506, 17 Lexington Avenue, New York, NY 10010} 
\address{Baruch College, The City University of New York, Natural Science Department, A506, 17 Lexington Avenue, New York, NY 10010}
\begin{abstract}

In earlier papers\cite{1985,1988,1990,1991} G\"{u}rsey \textit {et al.} showed development of a bilocal baryon-meson field from two quark-antiquark fields. 
The Hamiltonian in the case of vanishing quark masses was shown to have a very good agreement with experiments \cite{1990}. The theory for vanishing mass was solved using Confluent Hypergeometric functions \cite{1991}. 

In this paper I construct the normalized wave function for the spin-free Hamiltonian with light quark masses (only up to the first order of the mass of quark). I develop the new kind of special function theory in mathematics that generalize all existing theories of Confluent Hypergeometric types. 
I call it the 'Grand Confluent Hypergeometric (GCH) Function.' My solution produces previously unknown extra "hidden" radial quantum numbers relevant for description of supersymmetry and for generating new mass formulas. 

This paper is 1st out of 10 in series ``Special functions and three term recurrence formula (3TRF)''. See section 6 for all the papers in the series.  The next paper in the series describes generalization of three term recurrence relation in linear ordinary differential equations and its applications\cite{Chou2012b}.
 
\end{abstract}

\begin{keyword}
Supersymmetry; Semi-relativistic quark model; mass formula; Three term recurrence relation; Orthogonal relations

\PACS 02.30.Hq \sep 11.30.Pb \sep 12.40.Yx \sep 14.40.-n
\end{keyword}
                                      
\end{frontmatter}

\section{\label{sec:level1}Introduction} 
 
In 1974, Wilson showed how the string theory meet with an approximation to strongly interacting regime of QCD by the lattice gauge theory having a computable strong-coupling limit to QCD.\cite{Wils1974}
In 1975, Eguchi depicted that two quarks squeeze together and formed a bilocal linear structure with one quark at the end and a diquark at the other end through the string approximation with high rotational excitation.\cite{Eguc1975} A year later, Johnson and Thorn following the bag model for a baryon structure showed to QCD \cite{Chod1974.a,Chod1974.b} that the elongated bag model whose structure is controlled by tubes of color flux lines, stretched in a rotationally excited baryon, has an nearly linear Regge trajectory. And they calculated the universal Regge slope $\alpha =\frac{1}{4b}= 0.88 (GeV)^{-2}$.\cite{John1976}

G\"{u}rsey noticed that rotationally excited baryon and meson conduct like a elongated two-body system hold by tubes of color flux lines behaving as a scalar linear potential: (1) for meson, a quark is separated from an antiquark, (2) for baryon, a quark is separated from an diquark. And at large separation, since QCD forces are flavor independent and the confining part of the QCD scalar potential is spin independent, the $q-qq$ force is equal to the $q-\bar{q}$ force in the SU(3) case

According to G\"{u}rsey \textit{et al.} (1991 \cite{1991}), ``We derive an effective Hamiltonian of the relativistic quark model. In the limit of zero quark masses, we obtain linear Regge trajectories for mesons. Based on the diquark-antiquark symmetry, we show that the Regge trajectories of baryons and mesons are parallel at high angular moments. We discuss the breaking of the hadronic supersymmetry and obtain a mass relation of mesons and baryons.'' 
Following their analysis, the Hamiltonian in the case of vanishing quark masses was shown to have a very good agreement with experiments \cite{1990}. Since they neglect mass of quark in a supersymmetric differential equation \cite{1991}, the power series expansion in closed forms consists of two term recursion relation. The theory for vanishing mass was solved using Confluent Hypergeometric function. 

In this paper I include small mass of quark in their supersymmetric differential equation and its power series expansion in closed forms consists of three term recurrence relation. I develop a new kind of special function that generalizes the Confluent Hypergeometric series that I call Grand Confluent Hypergeometric (GCH) function.

Recently Heun function started to appear in theoretical modern physics. For example the Heun functions come out in the hydrogen-molecule ion\cite{Wils1928}, in the Schr$\ddot{\mbox{o}}$dinger equation with doubly anharmonic potential\cite{Ronv1995} (it's solution is the confluent forms of Heun function), in the Stark effect\cite{Epst1926}, in perturbations of the Kerr metric\cite{Teuk1973,Leav1985,Bati2006,Bati2007,Bati2010}, in crystalline materials\cite{Slavy2000}, in Collogero-Moser-Sutherland systems\cite{Take2003}, etc., just to mention a few.\cite{Birk2007,Hortacsu:2011rr,Suzu1998,Suzu1999}

The GCH ordinary differential equation is of Fuchsian types with the one regular and one irregular singularities in (\ref{eq:39}). In contrast, Heun equation of Fuchsian types has four regular singular points.\cite{Heun1889,Ronv1995} Heun equation has the four kind of confluent forms: (a) Confluent Heun (two regular and one irregular singularities), (b) Doubly confluent Heun (two irregular singularities), (c) Biconfluent Heun (one regular and one irregular singularities), (d) Triconfluent Heun equations (one irregular singularity). Biconfluent Heun equation is derived from the GCH equation by changing all coefficients $\mu = 1$ and $\varepsilon \omega =-q$. \cite{Chou2012i,Chou2012j}
In this paper I will show how Confluent Hypergeometric function is related to Grand Confluent Hypergoemetric function analytically.

Due to its complex mathematical calculation in three term recurrence relation of their linear ordinary differential equation \cite{Hortacsu:2011rr}, I construct the analytic solution of their supersymmetric differential equation only up to the first order of the extremely small mass of quark. More than second order of the mass of quark is negligible in this paper. 

As the mass of quark is negligible in their effective Hamiltonian of the relativistic quark model, its differential equation turns to be Confluent Hypergeometric differential equation. As we all know, there is only one eigenvalue and it has infinite eigennumbers which is called radial quantum number. For another example a hydrogen like atom wave function, only has one eigenvalue and has infinite eigennumbers. 

In contrast infinite eigenvalues is arisen in their supersymmetric differential equation as the small mass of quark is included in their Hamiltonian. Each eigenvalue has infinite eigennumbers \cite{Chou2012i,Chou2012j}. 
The concept of its eigenvalues gives rise to an extra degree of a quantum number I designate as ``$i^{th}$ kind of hidden radial quantum number'' that will be expressed below. 
This makes it especially applicable to supersymmetric theories having three term recurrence relation in the power series expansion of their differential equations in nature. 
As we see in Regge trajectory plot of angular momentum vs. square of mass (J vs. $m^2$), there are many linearly increasing lines with same slopes including bunch of eigenvalues corresponding to fermions and bosons. It is not clear what the meaning of many eigenvalues are: more details are explained in section 4.3 \cite{Chou2012i}. 

In section 2, I consider asymptotic behaviors of GCH differential equation including only up to the first order of $\frac{\varepsilon }{2}$ ($\varepsilon$ corresponds to the mass of quark). In section 3 and 4, I construct the power series expansion in closed forms and the integral representation of GCH function with including the first order of $\frac{\varepsilon }{2}$ for a polynomial and an infinite series. Also I derive the generating function and orthogonal relation of GCH polynomial with including the first order of $\frac{\varepsilon }{2}$. 
\subsection{Analytic solution neglecting the mass of quark in effective Hamiltonian of the relativistic quark model} 
Following Lichitenberg \textit{et al.}\cite{Lich1982}, G\"{u}rsey and collaborators wrote the spin free Hamiltonian involving only scalar potential for the $q-\bar{q}$ system:\cite{1985,1988,1990,1991}, namely           
\begin{equation} 
H^2 = 4\left[(m+ \frac{1}{2}V_s)^2 + P_r^2 + \frac{l(l+1)}{r^2}\right] 
\label{eq:1}
\end{equation}
where $P_r^2 = - \frac{\partial ^2}{\partial r^2} - \frac{2}{r} \frac{\partial}{\partial r} $, $m$= mass of quark, $V_s =br$ is the scalar potential with $r$ being the distance between the constituents in the bilocal linear system where $b$= real positive, and $l$= angular momentum quantum number. G\"{u}rsey \textit{et al.} assume that $m$ can be neglected because m is extremely small for u and d quarks \cite{1991}. (\ref{eq:1}) turns to be
\begin{equation}
H^2 \approx  4\left[\frac{1}{4}b^2 r^2 + P_r^2 + \frac{l(l+1)}{r^2}\right]
\label{eq:2}
\end{equation}
Its normalized wave function by using orthogonal relation is
\begin{equation}
\psi(r,\theta ,\phi ) \approx \sqrt{\frac{b^{\gamma }}{2^{\gamma -1} \Gamma (|\alpha _{0}|+1)\Gamma (|\alpha _{0}|+\gamma ) }}\,r^l e^{-\frac{1}{4}br^2}F_{|\alpha_0|}\left(\gamma = l+3/2 ;z\right) Y_l^{m^{\star}}(\theta ,\phi )
\label{eq:10}
\end{equation}
where
\begin{subequations}
\begin{equation}
\alpha_0 = 1 - n_0 = - \frac{1}{2b} \left( E^2/4 - \left(l+ 3/2 \right) b \right)=0,-1,-2,-3,\cdots
\label{eq:3a}
\end{equation}
$n_0 = 1,2,3,\cdots$ which is primary radial quantum number.
\begin{equation}
F_{|\alpha _0|}(\gamma = l+3/2;z) = \frac{\Gamma (|\alpha _{0}|+\gamma )}{\Gamma (\gamma )}\sum_{n=0}^{|\alpha _{0}|} \frac{(-|\alpha _{0}|)_n}{n!(\gamma )_n} z^n
\label{eq:3b}
\end{equation}
\end{subequations}
(\ref{eq:3b}) is the first kind of the independent solution of Confluent Hypergeometric function. And its eigenvalue is 
\begin{equation}
E^2 = 4b\left( 2|\alpha_0| + l+3/2 \right) = 4b\left( 2 n_0 + l-1/2 \right)
\label{eq:4}
\end{equation} 
Furthermore, Iachello \textit {et al.} obtained a similar mass formula for the stringlike properties of mesons with two fixed parameters, based on the spectrum-generating algebra $\mathit{G}$ in 1991.\cite{Iach1991} 
\subsection{Analytic solution including only up to the first order of the mass of quark in effective Hamiltonian of the relativistic quark model} 
In this paper I consider normalized wave function including the small mass $m$: only up to 1st order of $m$ terms.
When wave function $\psi (r) = e^{-\frac{b}{4}\left(r+\frac{2m}{b}\right)^2} r^l y(r)Y_l^{m^{\star}}(\theta ,\phi)$ acts on both sides of (\ref{eq:1}), it becomes
\begin{equation}
r\frac{\partial^2{y}}{\partial{r}^2} + \left( - b r^2 -2m r +2(l+1)\right) \frac{\partial{y}}{\partial{r}} + \left( Q r  -2m(l+1)\right) y = 0
\label{eq:11}
\end{equation}
where
\begin{equation}
Q= E^2/4- b\left(l+3/2\right)
\label{eq:12}
\end{equation}
By using the function $y(r)$ as Frobinous series in (\ref{eq:11}), I obtain two indicial roots which are $\lambda _1 = 0 $ and $\lambda _2 = -2l-1$. 
Recurrence formula for all $n$ is
\begin{equation}
K_n = A_n +\frac{B_n}{K_{n-1}} 
\begin{cases} K_n= \frac{c_{n+1}}{c_{n}} \cr
K_{n-1}= \frac{c_n}{c_{n-1}}\cr
A_n= \frac{2m(n+\lambda +l+1)}{(n+\lambda +1)(n+\lambda +2(l+1))} \cr
B_n= \frac{-Q+b(n+\lambda -1)}{(n+\lambda +1)(n+\lambda +2(l+1))}
\end{cases}
\label{eq:14}
\end{equation}
where $c_1=A_0 c_0 = m c_0$ and $n\geq 1$.

Let's investigate function $y(r)$ as $n$ and $r$ go to infinity.
As $n\gg 1$ (for sufficiently large), (\ref{eq:14}) is
\begin{equation}
\lim_{n\gg 1}K_n = \frac{2m}{n}+ \frac{b/n}{{\displaystyle \lim_{n\gg 1}K_{n-1}}} 
\label{eq:15}
\end{equation}
The first term of RHS in (\ref{eq:15}) is negligible, since mass $m$ is extremely small and $n$ is too large, respectively. Then, (\ref{eq:15}) is approximately equal to 
\begin{equation}
\lim_{n\gg 1}K_n \approx \frac{b/n}{{\displaystyle\lim_{n\gg 1}K_{n-1}}}
\label{eq:16}
\end{equation}  
Classify $c_n$ to its even and odd parts from (\ref{eq:16}) by using $K_n= \frac{c_{n+1}}{c_{n}}$ and $K_{n-1}= \frac{c_n}{c_{n-1}}$.
\begin{equation}
\begin{tabular}{  l  l }
  \vspace{2 mm}
  \large{$c_0$} &\hspace{1cm} \large{$c_1$} = $m c_0$ \\
  \vspace{2 mm}
  \large{$c_2 = b c_0$} &\hspace{1cm} \large{$c_3 = \frac{b}{2}m c_0$} \\
  \vspace{2 mm}
  \large{$c_4 = \frac{1}{1\cdot 3}b^2 c_0$} &\hspace{1cm}  \large{$c_5 = \frac{b^2}{2^2 2!}m c_0$}\\
  \vspace{2 mm}
  \large{$c_6 = \frac{1}{1\cdot 3\cdot 5}b^3 c_0$} &\hspace{1cm}  \large{$c_7 = \frac{b^3}{2^3 3!}m c_0$}\\
  \vspace{2 mm}
  \large{$c_8 = \frac{1}{1\cdot 3\cdot 5\cdot 7}b^4 c_0$} &\hspace{1cm} \large{$c_9 = \frac{b^4}{2^4 4!}m c_0$} \\
  \vspace{2 mm}
  \large{$c_{10} = \frac{1}{1\cdot 3\cdot 5\cdot 7\cdot 9}b^5 c_0$} &\hspace{1cm} \large{$c_{11} = \frac{b^5}{2^5 5!}m c_0$} \\
  
 \hspace{2 mm} \large{\vdots} & \hspace{1.5 cm}\large{\vdots} \\
\end{tabular}
\label{eq:17}
\end{equation}
From (\ref{eq:17}) suggesting $c_0=1$ , the function $y(r)$ approximately is 
\begin{equation}
\lim_{n\gg 1}y(r) \approx \sum_n \frac{2^{2n} n!}{(2n)!} \bigg(\frac{1}{2}b r^2\bigg)^{n} + m r \sum_n \frac{1}{n!} \bigg(\frac{1}{2}b r^2\bigg)^{n} 
 > (1+m r)e^{\frac{1}{2}b r^2}
\label{eq:18}
\end{equation}
It is unacceptable that wave function $\psi (r,\theta ,\phi )$ is divergent as $r$ goes to infinity from the quantum mechanical point of view. As $r$ is extremely large value, the big polynomial of degree $n$ will take a dominant position. Substitute (\ref{eq:18}) into the wave function which gives $\psi (r,\theta ,\phi ) = N e^{-\frac{b}{4}\left(r+\frac{2m}{b}\right)^2} r^l y(r) Y_l^{m^{\star}}(\theta ,\phi)$ where $N$ is normalized constant.
\begin{equation}
\lim_{\substack{n\gg 1\\r\to \infty}} \psi (r,\theta ,\phi )  >  \lim_{r\to \infty} N ( 1 +m r) r^l e^{\frac{1}{4}b r^2} Y_l^{m^{\star}}(\theta ,\phi) \rightarrow \infty   
\label{eq:20}  
\end{equation}
Even if the mass $m$ is extremely small, the wave function $\psi (r,\theta ,\phi )$ will blows up as $r$ $\rightarrow \infty $. 
All wave functions must to go to zero as $r$ goes to infinity from a quantum mechanical perspective. The first and second term of $y(r)$ must also be terminated to become a polynomial of degree $n$ in this case. As we see in (\ref{eq:18}), the first term indicates  even term of $c_n$, and the second term of it has odd term of $c_n$.
Now, let's try to define the first kind of independent solution as $\lambda _1 = 0$. 

As $\lambda _1 =0$,
\begin{subequations}
\begin{equation}
A_n|_{\lambda =0}= \frac{2(n+l+1)m}{(n+1)(n+2(l+1))}
\label{eq:21a}
\end{equation}
\begin{equation}
B_n|_{\lambda =0}= \frac{-Q+b(n-1)}{(n+1)(n+2(l+1))}
\label{eq:21b}
\end{equation}
\end{subequations}
I define $B_{i,j,k,l}$ refering to $B_i B_j B_k B_l$. Classify $c_n$ to its even and odd parts up to the first order of small mass $m$ from (\ref{eq:14}).
\begin{eqnarray}
\begin{tabular}{  l  l }
  \vspace{2 mm}
  $c_0$ & $c_1= c_0 A_0 $ \\
  \vspace{2 mm}
  $c_2 = c_0 B_1$ &$c_3 = c_0(A_0 B_2 +A_2 B_1)$ \\
  \vspace{2 mm}
  $c_4 = c_0 B_{1,3}$ & $c_5 = c_0 (A_0 B_{2,4} +A_2 B_{1,4}+A_4 B_{1,3})$\\
  \vspace{2 mm}
  $c_6 = c_0 B_{1,3,5}$ &  $c_7 = c_0(A_0 B_{2,4,6} +A_2 B_{1,4,6}+A_4 B_{1,3,6}+A_6 B_{1,3,5})$\\
  \vspace{2 mm}
  $c_8 = c_0 B_{1,3,5,7}$ & $c_9 = c_0(A_0 B_{2,4,6,8} +A_2 B_{1,4,6,8}+A_4 B_{1,3,6,8}$\\
  \vspace{2 mm}
  & \hspace{0.7 cm}$+A_6 B_{1,3,5,8}+A_8 B_{1,3,5,7})$\\
  \vspace{2 mm}
  $c_{10} = c_0 B_{1,3,5,7,9}$ & $c_{11} = c_0(A_0 B_{2,4,6,8,10} +A_2 B_{1,4,6,8,10}+A_4 B_{1,3,6,8,10}$\\                                  
  \vspace{2 mm}
   &\hspace{0.7 cm} $+A_6 B_{1,3,5,8,10}+ A_8 B_{1,3,5,7,10}+A_{10} B_{1,3,5,7,9})$ \\                                         
 \hspace{2 mm} \vdots & \hspace{1.5 cm} \vdots  \\
\end{tabular}
\label{eq:22}
\end{eqnarray}
As I describe $y(r)$ as a power series by using (\ref{eq:22}),
\begin{eqnarray}
y(r) &=& y(r)_{ domin.}+y(r)_{small} = \sum_{n=0}^{\infty } c_{2n} r^{2n} + \sum_{n=0}^{\infty }c_{2n+1} r^{2n+1} \nonumber\\ \! &=&\!   c_0 \bigg\{\! 1+\sum_{n=1}^{\infty }\prod_{k=0}^{n-1}B_{2k+1}r^{2n}\bigg\}\! + c_0 \bigg\{\! A_0 r + (A_0 B_2+A_2 B_1)r^3 \label{eq:23}\\
\!&+&\! \sum_{n=2}^{\infty }\bigg(\! A_0\prod _{p=0}^{n-1}B_{2p+2}+\sum_{j=1}^{n-1}\bigg(\! A_{2j}\prod _{p=j}^{n-1}B_{2p+2}\prod _{k=0}^{j-1}B_{2k+1}\bigg)\! + A_{2n}\prod _{p=0}^{n-1}B_{2p+1}\bigg) r^{2n+1}\bigg\} 
\nonumber
\end{eqnarray}
I choose one of $B_{2k+1}$ to be zero, where $k=0,1,2,3,\cdots$ in order to make a polynomial of degree $n$ in  $y(r)_{ domin.}$ of (\ref{eq:22}). In other words, I might choose $Q=2b(n_0-1)$ where $n_0=1,2,3,\cdots$ in (\ref{eq:21b}). Then $B_{2k+1}$ will be zero at certain value of $k$. However, as we see  $y(r)_{ small}$ in (\ref{eq:23}) (look at the odd term of $c_n$ in (\ref{eq:22})), it has combinations of $B_{2k+2}$ and $B_{2k+1}$ for every odd $c_n$ term. It means that one of each $B_{2k+2}$ and $B_{2k+1}$ terms must be zero at same time. In other words, $Q=2b(n_0-1)$ and $Q=b(2n_1-1)$ where $n_1=0,1,2,3,\cdots$ must be satisfied in this series simultaneously. I call $n_0$ as primary radial quantum number and $n_1$ as the first kind of hidden radial quantum number. Then $y(r)$ will be a polynomial degree of $n$. $y(r)$ consists of two terms which are $y(r)_{domin.}$ and $y(r)_{small}$ in (\ref{eq:23}). The dominant wave function $y(r)_{domin.}$ which does not include small mass $m$ must be terminated to become a polynomial of degree $n$ in this case. $y(r)_{small}$ is also polynomial and extremely small wave function because it includes $A_n|_{\lambda =0}= \frac{2(n+l+1)m}{(n+1)(n+2(l+1))}$ which has small mass $m$. Also (\ref{eq:12}) is equivalent to $2b(n_0-1)$ and $b(2n_1-1)$ at same time. And in this paper Pochhammer symbol $(x)_n$ is used to represent the rising factorial: $(x)_n = \frac{\Gamma (x+n)}{\Gamma (x)}$.

For simplicity plugging $c_0= \frac{(l+n_0-1/2)!}{(l+1/2)!}$ in (\ref{eq:23}), I obtain
\begin{eqnarray}
y(r) &=& \mathcal{QW}_{|\alpha _{0}|,|\alpha _{1}|}\Big(|\alpha _{0}|=n_0-1,|\alpha _{1}|=n_1-1,\gamma =l+3/2; z=\frac{1}{2}br^2\Big) \nonumber\\ 
&=& F_{|\alpha _0|}(\gamma ;z)+m r \sideset{_{}^{}}{_{|\alpha _{0}|}^{|\alpha _{1}|}}\prod \big(\gamma ;z\big) \hspace{2cm} \mathrm{only}\; \mathrm{if} \; |\alpha _{0}|\leq |\alpha _{1}|
\label{eq:24}
\end{eqnarray}
where,
\begin{subequations}
\begin{equation}
F_{|\alpha _0|}(\gamma ;z) = \frac{\Gamma (|\alpha _{0}|+\gamma )}{\Gamma (\gamma )}\sum_{n=0}^{|\alpha _{0}|} \frac{(-|\alpha _{0}|)_n}{n!(\gamma )_n} z^n
\label{eq:25a}
\end{equation}
\begin{eqnarray}
\sideset{_{}^{}}{_{|\alpha _{0}|}^{|\alpha _{1}|}}\prod \big(\gamma ;z\big) &=&  \frac{\Gamma (|\alpha _{0}|+\gamma )}{\Gamma (\gamma )}\sum_{n=0}^{|\alpha _{0}|} \frac{(-|\alpha _{0}|)_n}{n!(\gamma )_n} z^n \label{eq:25b}\\
&&\times \sum_{k=0}^{|\alpha _{1}|-n} \frac{\big(n+\frac{1}{2}(\gamma -\frac{1}{2})\big)\Gamma (n+\frac{1}{2})\Gamma (\gamma +n-\frac{1}{2})(n-|\alpha _{1}|)_k}{\Gamma (k+n+\frac{3}{2})\Gamma (k+n+\gamma +\frac{1}{2})} z^k \nonumber\\
&=& \frac{1}{2\pi i B(|\alpha _1|  +1,\frac{1}{2})} \mathcal{T}(s,t,p,u) \left( w_1 \partial_{w_1} +\frac{1}{2}(\gamma -1/2) \right) F_{|\alpha _0|}(\gamma ;w_1) 
\nonumber
\end{eqnarray}
\end{subequations}
And,
\begin{subequations}
\begin{equation}
B(|\alpha _1|  +1,\frac{1}{2}) = \frac{\Gamma (|\alpha _1|  +1)\Gamma (1/2)}{\Gamma (|\alpha _1|  +3/2)}
\label{eq:26a}
\end{equation}
\begin{equation}
w_1= \frac{zust}{(1-u)}(1-p^2)
\label{eq:26b}
\end{equation}
\end{subequations}
in the above, $\mathcal{T}(s,t,p,u)$ is the operator which acts on
\begin{equation}
\mathcal{T}(s,t,p,u)= \int_0^{\infty }ds \;s^{-1/2}(1+s)^{-(|\alpha _{1}|+3/2)}\int_0^1 dt \;t^{\gamma -3/2} \int_{-1}^1 dp \oint du \frac{e^{-\frac{zu(1-t)}{(1-u)}(1-p^2)}}{u^{|\alpha _{1}|+1}(1-u)}
\label{eq:27}
\end{equation}
We see in (\ref{eq:25a}), it is the first kind of confluent hypergeometric polynomial of degree $|\alpha _{0}|$. (\ref{eq:24}) denoted as $\mathcal{QW}_{|\alpha _{0}|,|\alpha _{1}|}\Big(|\alpha _{0}|=n_0-1,|\alpha _{1}|=n_1-1,\gamma =l+\frac{3}{2}; z=\frac{1}{2}br^2\Big)$ is the first kind of Grand Confluent Hypergeometric (GCH) polynomial of degree $|\alpha _{0}|$ and $|\alpha _{1}|$ with the first order $m$.

Also I obtain two eigenvalues which are $E_0^2= 4b\left\{l+2n_0-1/2\right\} $ and $E_1^2= 4b\left\{l+2n_1+1/2\right\} $.
The former is the primary radial eigenvalue. And the latter is the first kind of hidden radial eigenvalue.
As I let the small mass $m$ goes to zero in (\ref{eq:24}), its solution is same as the $1^{st}$ kind of confluent hypergeometric polynomial.
From (\ref{eq:24}), the wave function of it is
\begin{equation}
\psi(r,\theta ,\phi )_{n_0,n_1,l,m^{\star}} = \bar{N}  e^{-\frac{b}{4}\left(r+\frac{2m}{b}\right)^2} r^l \mathcal{QW}_{|\alpha _{0}|,|\alpha _{1}|}\Big(\gamma =l+\frac{3}{2}; z=\frac{1}{2}br^2\Big) Y_l^{m^{\star}}(\theta ,\phi)
\label{eq:29}
\end{equation}
By using orthogonal relation, normalized constant $\bar{N}$ is
\begin{eqnarray}
\bar{N} &=& \bigg [ \frac{2^{\gamma -1}}{b^{\gamma }}\Gamma (|\alpha _{0}|+1)\Gamma (|\alpha _{0}|+\gamma )\nonumber\\
&&- m \frac{(-1)^{|\alpha _{0}|}2^{\gamma -\frac{1}{2}}}{b^{\gamma +\frac{1}{2}}}\bigg\{ \frac{2 \Gamma(|\alpha _{0}|+\gamma -\frac{1}{2}) \Gamma(|\alpha _{0}|+\gamma +\frac{1}{2}) }{\Gamma (\gamma -\frac{1}{2})} \label{eq:30} \\
&&- \sum_{n=0}^{|\alpha _{0}|}\sum_{k=0}^{|\alpha _{1}|-n} \frac{(n+\frac{1}{2}(\gamma -\frac{1}{2})) \Gamma (|\alpha _{0}|+\gamma ) \Gamma (\gamma +n-\frac{1}{2})(-|\alpha _{0}|)_n (n-|\alpha _{1}|)_k}{\Gamma (\gamma )\Gamma (k+n-|\alpha _{0}|+\frac{1}{2})(\gamma )_n (n!)} \bigg\} \bigg] ^{- \frac{1}{2} }
\nonumber
\end{eqnarray}
where,
\begin{equation}
\left. 
\begin{array}{@{}r@{\quad}ccrr@{}}
|\alpha _{0}| = n_0 -1 \\
|\alpha _{1}|= n_1 -1           
\end{array} 
\right\} \textrm{only}\; \textrm{if}\; |\alpha _{0}|\leq |\alpha _{1}|
\label{eq:31}
\end{equation}
(\ref{eq:29}) is the $1^{st}$ kind of the normalized Grand confluent hypergeometric (GCH) wave function of degree $|\alpha _{0}|$ and $|\alpha _{1}|$.

As the small mass $m$ goes to zero in (\ref{eq:29}) and (\ref{eq:30}), it turns out
\begin{equation}
\lim_{m\rightarrow 0}\psi(r,\theta ,\phi )_{n_0,n_1,l,m^{\star}}  =  \sqrt{\frac{b^{\gamma }}{2^{\gamma -1} \Gamma (|\alpha _{0}|+1)\Gamma (|\alpha _{0}|+\gamma ) }}r^l e^{-\frac{1}{4}br^2}F_{|\alpha_0|}\left(\gamma ;z\right) Y_l^{m^{\star}}(\theta ,\phi )
\label{eq:32}
\end{equation}
(\ref{eq:32}) is equivalent to (\ref{eq:10}). 
There are two eigennumbers $n_0$ and $n_1$ in (\ref{eq:29}). The first eigennumber $n_0$, called primary radial quantum number, appears in zeroth order of $m$ term which is $y(r)_{domin.}$. And the second eigennumber $n_1$, called the first kind of hidden radial quantum number, appears in the first order of the $m$ term which is $y(r)_{small}$. As I neglect small mass $m$, the primary radial quantum number only starts to appear in the wave function. However when I include the small mass $m$, the second eigennumber which is $1^{st}$ kind of hidden radial quantum number is created. As we sees any other special functions such as Laguerre and Associated laguerre functions, Legendre and associated Legendre functions, hypergeometric function, Kummer function, etc, those functions only have one eigenvalue. 

In this new special function, there are two terms which are $0^{th}$ order of $m$ term such as $y(r)_{domin.}$ and $1^{st}$ order of $m$ term such as $y(r)_{small}$. Actually, higher order of $m$ terms do exist. But the mass of quark is extremely small. So I neglect the $m$ terms that are more than $2^{nd}$ order. I only count it up to a $1^{st}$ order of $m$ term. The full description of function $y(r)$ include all higher order of mass $m$ in the following way.
\begin{eqnarray}
y(r) &=& \sum_{i=0}^{\infty } m^{i} \sum_{n=0}^{\infty} C_{n,i} r^n \label{eq:33}\\
&=& \sum_{n=0}^{\infty} C_{n,0}\; r^n + m \sum_{n=0}^{\infty} C_{n,1} \;r^n +m^2 \sum_{n=0}^{\infty} C_{n,2} \;r^n + m^3 \sum_{n=0}^{\infty} C_{n,3} \;r^n + \cdots
\nonumber
\end{eqnarray} 
If the function $y(r)$ of (\ref{eq:33}) is infinite series, then the wave function $\psi (r,\theta ,\phi )$ will blow up as we see in (\ref{eq:20}). So all of each of summation in (\ref{eq:33}) must be a polynomial. Then, (\ref{eq:33}) become
\begin{eqnarray}
y(r) &=& \sum_{i=0}^{\infty } m^{i} \sum_{n=0}^{N_i} C_{n,i} r^n \label{eq:34}\\
&=& \sum_{n=0}^{N_0} C_{n,0}\; r^n + m \sum_{n=0}^{N_1} C_{n,1} \;r^n +m^2 \sum_{n=0}^{N_2} C_{n,2} \;r^n + m^3 \sum_{n=0}^{N_3} C_{n,3} \;r^n + \cdots
\nonumber
\end{eqnarray} 
On the above $N_i$ where $i=0,1,2,\cdots$ is the eigenvalue in order to make the sub-power series expansion as a polynomial.
Then, for example, all possible eigenvalues and eigennumbers of $m^{th}$ order term up to forth order of $m$ are
\begin{subequations}
\begin{equation}
\emph{m}^0 \;\textrm{order}\;\textrm{term} ; \left \{ \begin{array}{@{}r@{\quad}ccrr@{}} \vspace{2 mm}
E_0^2 & = & 4b\Big ( l+2n_0-\frac{1}{2}\Big ) & n_0 = 1,2,3,4,\cdots 
\end{array}
\right\} \label{eq:35a}
\end{equation}
\begin{equation}
\emph{m}^1 \;\textrm{order}\;\textrm{term} ; \left \{ \begin{array}{@{}r@{\quad}ccrr@{}} \vspace{2 mm}
E_0^2 & = & 4b\Big (l+2n_0-\frac{1}{2}\Big ) & n_0 = 1,2,3,4,\cdots \\ \vspace{2 mm}
E_1^2 & = & 4b\Big (l+2n_1+\frac{1}{2}\Big ) & n_1 = 1,2,3,4,\cdots 
\end{array}
\right\}\label{eq:35b}
\end{equation}
\begin{equation}
\emph{m}^2 \;\textrm{order}\;\textrm{term} ; \left \{ \begin{array}{@{}r@{\quad}ccrr@{}} \vspace{2 mm}
E_0^2 & = & 4b\Big (l+2n_0-\frac{1}{2}\Big ) & n_0 = 1,2,3,4,\cdots \\ \vspace{2 mm}
E_1^2 & = & 4b\Big (l+2n_1+\frac{1}{2}\Big ) & n_1 = 1,2,3,4,\cdots \\ \vspace{2 mm}
E_2^2 & = & 4b\Big (l+2n_2+\frac{3}{2}\Big ) & n_2 = 1,2,3,4,\cdots 
\end{array}
\right\} \label{eq:35c}
\end{equation}
\begin{equation}
\emph{m}^3 \;\textrm{order}\;\textrm{term} ; \left \{ \begin{array}{@{}r@{\quad}ccrr@{}} \vspace{2 mm}
E_0^2 & = & 4b\Big (l+2n_0-\frac{1}{2}\Big ) & n_0 = 1,2,3,4,\cdots \\ \vspace{2 mm}
E_1^2 & = & 4b\Big (l+2n_1+\frac{1}{2}\Big ) & n_1 = 1,2,3,4,\cdots \\ \vspace{2 mm}
E_2^2 & = & 4b\Big (l+2n_2+\frac{3}{2}\Big ) & n_2 = 1,2,3,4,\cdots \\ \vspace{2 mm}
E_3^2 & = & 4b\Big (l+2n_3+\frac{5}{2}\Big ) & n_3 = 1,2,3,4,\cdots
\end{array}
\right\} \label{eq:35d}
\end{equation}
\begin{equation}
\emph{m}^4 \;\textrm{order}\;\textrm{term} ; \left \{ \begin{array}{@{}r@{\quad}ccrr@{}} \vspace{2 mm}
E_0^2 & = & 4b\Big (l+2n_0-\frac{1}{2}\Big ) & n_0 = 1,2,3,4,\cdots \\ \vspace{2 mm}
E_1^2 & = & 4b\Big (l+2n_1+\frac{1}{2}\Big ) & n_1 = 1,2,3,4,\cdots \\ \vspace{2 mm}
E_2^2 & = & 4b\Big (l+2n_2+\frac{3}{2}\Big ) & n_2 = 1,2,3,4,\cdots \\ \vspace{2 mm}
E_3^2 & = & 4b\Big (l+2n_3+\frac{5}{2}\Big ) & n_3 = 1,2,3,4,\cdots \\ \vspace{2 mm}
E_4^2 & = & 4b\Big (l+2n_4+\frac{7}{2}\Big ) & n_4 = 1,2,3,4,\cdots
\end{array}
\right\} \label{eq:35e}
\end{equation}
\end{subequations} 
where
\begin{equation}
\begin{cases} n_i\leq n_j \;\textrm{where}\; i\leq j \;\textrm{and}\; i,j=0,1,2,\cdots \cr
E_0^2 = \textrm{primary}\; \textrm{radial} \;\textrm{eigenvalue} \cr
E_i^2 = \textrm{$i^{th}$}\;\textrm{type}\;\textrm{hidden} \;\textrm{radial}\;\textrm{eigenvalue} \cr
n_0 = \textrm{primary}\; \textrm{radial} \;\textrm{quantum}\;\textrm{number} \cr
n_i = \textrm{$i^{th}$}\;\textrm{type}\;\textrm{hidden} \;\textrm{radial}\;\textrm{quantum}\;\textrm{number} 
\end{cases}
\label{eq:36}
\end{equation}
For the high rotational excited bound state in the bag model, G\"{u}rsey \textit{et al.} showed that mesonic ($q-\bar{q}$ sysytem) and baryonic ($q-qq$ sysytem) trajectories are approximately parallel to each other with the universal Regge slope $\alpha =\frac{1}{4b}= 0.9 (GeV)^{-2}$. They also demonstrated why trajectories of these hadronic constituents are approximately linear with high rotational excitations. Even though they explain a phenomenological QCD model of rotationally excited $q-\bar{q}$ and $q-D$ supersymmetries in the bag model which is behind Miyazawa's SU(6/21) scheme analytically, they do not show why the mesonic and baryonic trajectories have the same separation nearly. They neglect the mass of quark in their supersymmetric Hamiltonian due to its complex computations. From this assumption, they only obtain one eigenvalue (mass formula) of meson and baryon for high rotational excitation in (\ref{eq:4}).

However, as I include the mass of quark in their semi-relativistic wave equation for the hadronic wavefunction, I obtain the infinity eigenvalues (mass formula for meson for high rotational excitation): we see $i^{th}$ term of $m$ has (i+1) different eigenvalues from (\ref{eq:35a})-(\ref{eq:35e}). Actually we don't have to think about eigenfunctions of all higher order of $m^{th}$ term, since the mass of quark creates extremely small vibrations and variation because of small mass $m$. So zeroth order of $m$ and 1st order of $m$ terms are sufficient. However, I obtain many eigenvalues whenever the order of $m^{th}$ term increases. 
These are reasons why there are many linearly increasing lines with same slopes and separations (intercepts) including bunch of eigenvalues corresponding to fermions and bosons. From (\ref{eq:35a})-(\ref{eq:36}), we can obtain quiet exact mass formulas of many hadronic particles. More detail about Grand Confluent Hypergeometric (GCH) function of all higher order of $m^{th}$ term and its eigenvalues are explained analytically.\cite{Chou2012i,Chou2012j}  

According to Lichitenberg \textit{et al.}\cite{Lich1982}, the full description of semi-relativistic Hamiltonian in (\ref{eq:1}) is given by 
\begin{equation}
H \rightarrow H - V_c  \nonumber
\end{equation}
where
\begin{equation}
V_c = -\frac{4}{3}\frac{\alpha _s}{r} +k \frac{\bf{s_1}\cdot \bf{s_2}}{m_1 m_2} \nonumber
\end{equation}
The first term of RHS in vector potential $V_c$ is the fourth component of a vector potential like a Coulomb potential. $\frac{4}{3}$ is the color factor. $\alpha _s$ is the strong coupling constant. The second term of RHS in $V_c$ is the hyperfine structure correction due to gluon exchange where $k = |\psi _{12}(0)|^2$. In this paper, I assume that $V_c$ is negligible because quarks behave as free particles and quark-quark potential is almost spin independent for a large separation. In future paper, I will construct mass formula of the meson $(q-\bar{q})$ including $V_c$ for a short range force. 

\section{\label{sec:level2}Asymptotic Behavior of grand confluent hypergeometric function}
Now let's generalize this new special function. Suppose that there is a second order differential equation which is
\begin{equation}
x^2 y^{''}(x) + a_0 x y^{'}(x) +(a_1 x^4+ b_1 x^3 +c_1 x^2+ d_1)y(x)=0  \label{eq:37} 
\end{equation}
where $a_0, a_1, b_1, c_1, d_1 = \Re$ and $ 0\leq x\leq \infty$.
(\ref{eq:37}) is equivalent to (\ref{eq:1}). All coefficients in the above are correspondent to the following way. 
\begin{equation}
\begin{split}
& a_0 \longleftrightarrow   2 \\ & a_1 \longleftrightarrow  - \frac{b^2}{4} \\ & b_1 \longleftrightarrow  -m b \\
& c_1 \longleftrightarrow  \Big(\frac{E^2}{4} - m^2\Big) \\ & d_1 \longleftrightarrow -l(l+1) \\ & x \longrightarrow r
\end{split}
\label{eq:38}   
\end{equation}
When a function 
$y(x) = e^{\frac{i}{2}\sqrt{a_1}\big(x+\frac{b_1}{2 a_1}\big)^2} x^{\frac{1}{2}\big(-(a_0-1)\pm \sqrt{(a_0-1)^2 -4 d_1}\big)} g(x)$ 
acts on (\ref{eq:37}), I have
\begin{equation}
x g^{''}(x) + (\mu x^2 + \varepsilon x + \nu ) g^{'}(x) + (\Omega x + \varepsilon \omega  )g(x) = 0
\label{eq:39}
\end{equation}
where,
\begin{equation}
\begin{split}
& \mu  =   2 i \sqrt{a_1} \\ & \varepsilon  = \frac{i b_1}{\sqrt{a_1}} \\ & \nu  =  1\pm \sqrt{(a_0-1)^2 -4 d_1} \\
& \Omega  = c_1 - \frac{b_1^2}{4 a_1} + i \sqrt{a_1}\big( 2 \pm \sqrt{(a_0-1)^2 -4 d_1} \big) \\ & \omega  = \frac{1}{2}\big( 1 \pm \sqrt{(a_0-1)^2 -4 d_1} \big) 
\end{split}  
\label{eq:40} 
\end{equation}
(\ref{eq:39}) is equivalent to (\ref{eq:11}). In my definition, (\ref{eq:39}) is the Grand Confluent Hypergeometric (GCH) differential equation. All coefficients in the above are correspondent to to the following: 
\begin{equation}
\begin{split}
& \mu  \longleftrightarrow   -b \\ & \varepsilon  \longleftrightarrow  -2 m \\ & \nu  \longleftrightarrow  2(l+1) \\
& \Omega \longleftrightarrow  Q = \frac{E^2}{4} - b\Big( l+\frac{3}{2} \Big)\\ & \omega  \longleftrightarrow (l+1) \\ & x \longrightarrow r
\end{split} 
\label{eq:41} 
\end{equation}
First I suggest that $|\frac{\varepsilon}{2}| = |\frac{i b_1}{2\sqrt{a_1}}| \ll 1 $. By using the function $g(x)$ as Frobinous series in (\ref{eq:39}), I have two indicial roots, $\lambda _1 = 0 $ and $\lambda _2 = 1-\nu $. And, recurrence formula for all $n$ is
\begin{equation}
K_n = A_n +\frac{B_n}{K_{n-1}} \hspace{.5cm}; n\geq 1
\begin{cases} K_n= \frac{c_{n+1}}{c_{n}} \cr
K_{n-1}= \frac{c_n}{c_{n-1}}\cr
A_n= -\frac{\varepsilon (n+\lambda +\omega )}{(n+\lambda +1)(n+\lambda +\nu )} \cr
B_n= -\frac{\Omega +\mu (n+\lambda -1)}{(n+\lambda +1)(n+\lambda +\nu )} \cr
\frac{c_1}{c_0} = -\frac{\varepsilon}{2} 
\end{cases}
\label{eq:43} 
\end{equation}
Let's test for the convergence of the function $g(x)$. As $n$ goes to infinity, recurrence formula is approximately equal to
\begin{equation}
c_{n+1} \simeq -\frac{\mu }{n} c_{n-1}
\label{eq:44} 
\end{equation} 
Now I can describe $g(x)$ as power series by substituting (\ref{eq:44}) into Frobinous series. For simplicity, I suggest $c_0=1$ 
\begin{eqnarray}
\lim_{n\gg 1}g(x) &\simeq &  \sum_n c_{2n} x^{2n} + \sum_{n} c_{2n+1} x^{2n+1}  \nonumber\\
&=& \sum_n \frac{(-\frac{1}{2})!}{(n-\frac{1}{2})!} \bigg(-\frac{1}{2}\mu  x^2\bigg)^{n}-\frac{1}{2}\varepsilon x  e^{-\frac{1}{2}\mu  x^2} \nonumber\\
&=& 1 + \sqrt{\pi }e^{(-\frac{1}{2}\mu  x^2)}  \sqrt{-\frac{1}{2}\mu  x^2}\; \mbox{Erf }\left(\sqrt{-\frac{1}{2}\mu  x^2}\right) -\frac{1}{2}\varepsilon x  e^{-\frac{1}{2}\mu  x^2} \nonumber\\
& >& (1-\frac{1}{2}\varepsilon x )e^{-\frac{1}{2}\mu  x^2}
\label{eq:45} 
\end{eqnarray}
In the above, $\displaystyle{\mbox{Erf }\left(\sqrt{-\frac{1}{2}\mu  x^2}\right)}$ is error function. Substitute (\ref{eq:45})  into $y(x)$.
\begin{eqnarray}
\lim_{n\gg 1}y(x) &\simeq & x^{\frac{1}{2}\big(-(a_0-1)\pm \sqrt{(a_0-1)^2 -4 d_1}\big)} \bigg(  e^{\frac{i}{2}\sqrt{a_1}x^2} -\frac{i b_1}{2 \sqrt{a_1}} x e^{-\frac{i}{2}\sqrt{a_1}x^2} \nonumber\\
&&+ (i)^{3/2}\sqrt{\pi} (a_1)^{1/4} x e^{-\frac{i}{2}\sqrt{a_1}x^2} \mbox{Erf }\left((i)^{3/2}(a_1)^{1/4}x\right)\bigg)\nonumber\\
& > & e^{-\frac{i}{2}\sqrt{a_1}x^2} x^{\frac{1}{2}\big(-(a_0-1)\pm \sqrt{(a_0-1)^2 -4 d_1}\big)}\Big(1-\frac{i b_1}{2 \sqrt{a_1}} x \Big)
\label{eq:46} 
\end{eqnarray}

\subsection{As $a_1= \mbox{real} \;\mbox{positive}$}

(a) If $\frac{1}{2}\big(-(a_0-1)\pm \sqrt{(a_0-1)^2 -4 d_1}\big) < -1$,

As $n\gg 1 $ and $x$ $ \rightarrow 0 $ in (\ref{eq:46}),
\begin{equation}
\lim_{\substack{n\gg 1\\x\to 0}}y(x) > x^{\frac{1}{2}\big(-(a_0-1)\pm \sqrt{(a_0-1)^2 -4 d_1}\big)}\Big(1-\frac{i b_1}{2 \sqrt{a_1}} x \Big) \rightarrow \infty  
\label{eq:47} 
\end{equation}
As $x\rightarrow \infty $ and $a_1\gg 1$ in $\mbox{Erf }\left((i)^{3/2}(a_1)^{1/4}x\right)$, the imaginary part of an error function oscillates around zero. And the real part of an error function oscillates at around $-1$. As $x$ goes to $\infty $ in (\ref{eq:46}) 
\begin{eqnarray}
\lim_{\substack{n\gg 1\\x\to \infty }}y(x) &\simeq&  x^{\frac{1}{2}\big(-(a_0-1)\pm \sqrt{(a_0-1)^2 -4 d_1}\big)} \bigg(  e^{\frac{i}{2}\sqrt{a_1}x^2} -\frac{i b_1}{2 \sqrt{a_1}} x e^{-\frac{i}{2}\sqrt{a_1}x^2}\nonumber\\ 
&+& (i)^{3/2}\sqrt{\pi} (a_1)^{1/4} x e^{-\frac{i}{2}\sqrt{a_1}x^2} \mbox{Erf }\left((i)^{3/2}(a_1)^{1/4}x\right)\bigg) \rightarrow 0
\label{eq:48} 
\end{eqnarray}

(b) If $\frac{1}{2}\big(-(a_0-1)\pm \sqrt{(a_0-1)^2 -4 d_1}\big) = -1$,

(\ref{eq:46}) turn out to be
\begin{eqnarray}
\lim_{n\gg 1}y(x) &\simeq & e^{-\frac{i}{2}\sqrt{a_1}x^2} \left\{\frac{1}{x} -\frac{i b_1}{2 \sqrt{a_1}} + (i)^{3/2}\sqrt{\pi} (a_1)^{1/4}  \mbox{Erf }\left((i)^{3/2}(a_1)^{1/4}x\right) \right\} \nonumber\\
&>& e^{-\frac{i}{2}\sqrt{a_1}x^2} \Big(\frac{1}{x}-\frac{i b_1}{2 \sqrt{a_1}} \Big) 
\label{eq:49} 
\end{eqnarray}
As $x\rightarrow 0 $ in (\ref{eq:49}),  it then yields
\begin{equation}
\lim_{\substack{n\gg 1\\x\to 0 }}y(x)> e^{-\frac{i}{2}\sqrt{a_1}x^2} \Big(\frac{1}{x}-\frac{i b_1}{2 \sqrt{a_1}} \Big) \rightarrow \infty 
\label{eq:50} 
\end{equation}
As $x$ goes to zero, the function $y(x)$ become divergent. Since $x\rightarrow \infty $ in (\ref{eq:49}), suggested by $|\frac{\varepsilon}{2}| = |\frac{i b_1}{2\sqrt{a_1}}| \ll 1 $, it is simply approximately
\begin{equation}
\lim_{\substack{n\gg 1\\x\to \infty}}y(x) \simeq  (i)^{3/2}\sqrt{\pi} (a_1)^{1/4}  e^{-\frac{i}{2}\sqrt{a_1}x^2}\mbox{Erf }\left((i)^{3/2}(a_1)^{1/4}x\right)
\label{eq:51} 
\end{equation}

(c) If $-1 < \frac{1}{2}\big(-(a_0-1)\pm \sqrt{(a_0-1)^2 -4 d_1}\big) < 0 $,

As $x$ goes to zero and $\infty $ in (\ref{eq:46}),
\begin{equation}
\lim_{\substack{n\gg 1\\x\to 0 \\ x\to \infty}}y(x) >  e^{-\frac{i}{2}\sqrt{a_1}x^2} x^{\frac{1}{2}\big(-(a_0-1)\pm \sqrt{(a_0-1)^2 -4 d_1}\big)}\Big(1-\frac{i b_1}{2 \sqrt{a_1}} x \Big) \rightarrow \infty 
\label{eq:52} 
\end{equation}

(d) If $ \frac{1}{2}\big(-(a_0-1)\pm \sqrt{(a_0-1)^2 -4 d_1}\big) = 0 $,

(\ref{eq:46}) turns out to be
\begin{eqnarray}
\lim_{n\gg 1}y(x) &\simeq &  e^{-\frac{i}{2}\sqrt{a_1}x^2} \left\{ 1-\frac{i b_1}{2 \sqrt{a_1}} x  + (i)^{3/2}\sqrt{\pi} (a_1)^{1/4} x \mbox{Erf }\left((i)^{3/2}(a_1)^{1/4}x\right) \right\}\nonumber\\
&>& e^{-\frac{i}{2}\sqrt{a_1}x^2} \Big(1 -\frac{i b_1}{2 \sqrt{a_1}} x \Big)
\label{eq:53} 
\end{eqnarray}
As $x$ goes zero and $\infty $ in (\ref{eq:53}),
\begin{equation}
\lim_{\substack{n\gg 1\\x\to 0}}y(x) \simeq  1 \hspace{1cm} \lim_{\substack{n\gg 1\\x\to \infty }}y(x) \rightarrow \infty 
\label{eq:54} 
\end{equation}
(e) If $ \frac{1}{2}\big(-(a_0-1)\pm \sqrt{(a_0-1)^2 -4 d_1}\big) > 0 $,

As $x$ goes to $\infty$ and zero in (\ref{eq:46}),
\begin{equation}
\lim_{\substack{n\gg 1\\x\to \infty }}y(x) > e^{-\frac{i}{2}\sqrt{a_1}x^2} x^{\frac{1}{2}\big(-(a_0-1)\pm \sqrt{(a_0-1)^2 -4 d_1}\big)}\Big(1-\frac{i b_1}{2 \sqrt{a_1}} x \Big) \rightarrow \infty 
\label{eq:56} 
\end{equation}
\begin{eqnarray}
\lim_{\substack{n\gg 1\\x\to 0 }}y(x) &\simeq&  x^{\frac{1}{2}\big(-(a_0-1)\pm \sqrt{(a_0-1)^2 -4 d_1}\big)} e^{-\frac{i}{2}\sqrt{a_1}x^2} \bigg( 1 -\frac{i b_1}{2 \sqrt{a_1}} x \nonumber\\
&+& (i)^{3/2}\sqrt{\pi} (a_1)^{1/4} x \mbox{Erf }\left((i)^{3/2}(a_1)^{1/4}x\right)\bigg) \rightarrow 0
\label{eq:57} 
\end{eqnarray}
\subsection{As $a_1= 0$}

(\ref{eq:46}) simply turns to be
\begin{equation}
\lim_{n\gg 1}y(x) \simeq  x^{\frac{1}{2}\big(-(a_0-1)\pm \sqrt{(a_0-1)^2 -4 d_1}\big)} \bigg( 1 -\frac{i b_1}{2 \sqrt{a_1}} x \bigg) 
\label{eq:58} 
\end{equation}
I suggest that $|\frac{\varepsilon}{2}| = |\frac{i b_1}{2\sqrt{a_1}}| \ll 1 $. Then as $a_1= 0$ in the second term of the bracket in (\ref{eq:58}), $|\frac{\varepsilon}{2}| = |\frac{i b_1}{2\sqrt{a_1}}| \rightarrow  \infty  $. The function $y(x)$ will be divergent no matter what the value of $x$ is. Therefore, there are no any independent solutions at all in the case of $a_1 = 0$.  

\subsection{As $a_1= \mbox{real} \;\mbox{negative}$}

Plug $a_1 = -|a_1|$ into (\ref{eq:46}).
\begin{eqnarray}
\lim_{n\gg 1}y(x) &\simeq &  x^{\frac{1}{2}\big(-(a_0-1)\pm \sqrt{(a_0-1)^2 -4 d_1}\big)}e^{\frac{1}{2}\sqrt{|a_1|}x^2}  
\nonumber\\
&&\times \bigg( 1- \frac{b_1}{2\sqrt{|a_1|}} x +\sqrt{\pi }(|a_1|)^{1/4} x \; \mbox{Erf }\left((|a_1|)^{1/4} x\right) \bigg) \nonumber\\
&>& e^{\frac{1}{2}\sqrt{|a_1|}x^2} x^{\frac{1}{2}\big(-(a_0-1)\pm \sqrt{(a_0-1)^2 -4 d_1}\big)}\Big(1-\frac{b_1}{2 \sqrt{|a_1|}} x \Big)
\label{eq:59} 
\end{eqnarray}
(a) If $ \frac{1}{2}\big(-(a_0-1)\pm \sqrt{(a_0-1)^2 -4 d_1}\big) < 0  $,

As $x$ goes to $0$ and $\infty$ in (\ref{eq:59}), the function $ y(x)$ is divergent.
\begin{equation}
\lim_{\substack{n\gg 1\\ x\to 0 \\x\to \infty }}y(x) >e^{\frac{1}{2}\sqrt{|a_1|}x^2} x^{\frac{1}{2}\big(-(a_0-1)\pm \sqrt{(a_0-1)^2 -4 d_1}\big)}\Big(1-\frac{b_1}{2 \sqrt{|a_1|}} x \Big) \rightarrow \infty 
\label{eq:60} 
\end{equation}
(b) If $ \frac{1}{2}\big(-(a_0-1)\pm \sqrt{(a_0-1)^2 -4 d_1}\big) = 0 $,

As $x$ goes to $0$ and $\infty $ in (\ref{eq:59}), 
\begin{equation}
\lim_{\substack{n\gg 1\\ x\to 0 }}y(x) \rightarrow 1 \hspace{1cm} \lim_{\substack{n\gg 1\\ x\to \infty  }}y(x) \rightarrow \infty 
\label{eq:61} 
\end{equation}
(c) If $ \frac{1}{2}\big(-(a_0-1)\pm \sqrt{(a_0-1)^2 -4 d_1}\big) > 0 $,

As $x$ goes to $0$ and $\infty $ in (\ref{eq:59}), 
\begin{equation}
\lim_{\substack{n\gg 1\\ x\to 0 }}y(x) \rightarrow 0 \hspace{1cm} \lim_{\substack{n\gg 1\\ x\to \infty  }}y(x) \rightarrow \infty 
\label{eq:63}
\end{equation}
\subsection{As  $g(x)$ is polynomial for $a_1= $ real negative }
I check all possible tests to determine whether an infinite series of function $y(x)$ converges or diverges on the above. Now let's test for convergence as the function $y(x)$ as $g(x)$ is polynomial for the case of $a_1 = $ real negative. Substitute $g(x) = {\displaystyle \sum_{n=0}^{N} C_n x^{n+\lambda}}$ into $y(x)$.
\begin{equation}
y(x)  \simeq  e^{-\frac{1}{2}\sqrt{|a_1|}x^2} x^{\frac{1}{2}\big(-(a_0-1)\pm \sqrt{(a_0-1)^2 -4 d_1}\big)} \sum_{n=0}^{N} C_n x^{n+\lambda}
\label{eq:65}
\end{equation}
(a) If $ \frac{1}{2}\big(-(a_0-1)\pm \sqrt{(a_0-1)^2 -4 d_1}\big) < 0  $,

As $x$ goes to $0$ and $\infty $ in (\ref{eq:65}), 
\begin{equation}
\lim_{\substack{n\gg 1\\ x\to 0 }}y(x) \rightarrow \infty \hspace{1cm} \lim_{\substack{n\gg 1\\ x\to \infty  }}y(x) \rightarrow 0 
\label{eq:66}
\end{equation}
(b) If $ \frac{1}{2}\big(-(a_0-1)\pm \sqrt{(a_0-1)^2 -4 d_1}\big) = 0  $,

As $x$ goes to $0$ and $\infty $ in (\ref{eq:65}), 
\begin{equation}
\lim_{\substack{n\gg 1\\ x\to 0 }}y(x) \rightarrow 1 \hspace{1cm} \lim_{\substack{n\gg 1\\ x\to \infty  }}y(x) \rightarrow 0 
\label{eq:68}
\end{equation}
c) If $ \frac{1}{2}\big(-(a_0-1)\pm \sqrt{(a_0-1)^2 -4 d_1}\big) > 0  $,

As $x$ goes to $0$ and $\infty $ in (\ref{eq:65}), 
\begin{equation}
\lim_{\substack{n\gg 1\\ x\to 0 \\x\to \infty }}y(x) \rightarrow 0
\label{eq:70}
\end{equation}
I choose boundary conditions of a function $g(x)$ for the polynomial in the following:
\begin{equation}
\begin{cases}\displaystyle{\lim_{x\to 0}}g(x)\rightarrow \mbox{convergent} \cr
\displaystyle{\lim_{x\to \infty }}g(x)\rightarrow 0
\end{cases}
\label{eq:71}
\end{equation}
and the necessary conditions of (\ref{eq:71}) is
\begin{equation}
\begin{cases} a_1 = \mbox{real} \; \mbox{negative} \cr
\frac{1}{2}\big(-(a_0-1)\pm \sqrt{(a_0-1)^2 -4 d_1}\big) \geq  0 \
\end{cases}
\label{eq:72}
\end{equation}
After we develop independent solutions for the polynomial, we can expand them as an infinite series in simple ways. More details about connections between a polynomial and an infinite series of the GCH equation are explained analytically.\cite{Chou2012i,Chou2012j} When we try to find the analytic solution of any differential equations, first we must consider what physical circumstance
and mathematical condition make solutions as a polynomial or an infinite series. 

\section{\label{sec:level3}Polynomial for $\nu$ =non-integer}
I consider the power series expansion in closed forms of the GCH polynomial only up to the $1^{st}$ order of $m$ terms, its integral forms, generating functions and orthogonal relations. As we know, there are two indicial roots which are $\lambda _1 =0$ and $\lambda _2 =1-\nu $. 
\subsection{ As $\lambda _1 = 0$ }
\subsubsection{Power series expansion in closed forms}
From (\ref{eq:43})
\begin{subequations}
\begin{equation}
A_n|_{\lambda =0}= - \frac{\varepsilon (n+\omega )}{(n+1)(n+\nu )}
\label{eq:73a}
\end{equation}
\begin{equation}
B_n|_{\lambda =0}= - \frac{\Omega +\mu (n-1)}{(n+1)(n+\nu )}
\label{eq:73b}
\end{equation}
\end{subequations}
Put $n=0$ in (\ref{eq:73a}). I obtain $A_0 = -\frac{\omega }{\nu }\varepsilon = -\frac{\varepsilon }{2} = \frac{c_1}{c_0}$. From (\ref{eq:43}), I have $K_0 =  \frac{c_1}{c_0} $ which is equal to $A_0$. Plug $n=1$ into recurrence formula.
\begin{equation}
K_1= A_1 + \frac{B_1}{K_0} = A_1 + \frac{B_1}{A_0}= A_1 - \frac{B_1}{(\varepsilon /2)}
\label{eq:74}
\end{equation} 
As we see in (\ref{eq:74}), $A_n$ includes the first order of $\frac{\varepsilon }{2}$ in which has an extremely small value. Then I argue that $|A_1| \ll |\frac{B_1}{A_0}| = \frac{B_1}{(\varepsilon /2)}$. (\ref{eq:74}) is approximatively equal to $ K_1\simeq \frac{B_1}{A_0} = - \frac{B_1}{(\varepsilon /2)}$. By using this process, I can simplify (\ref{eq:43}) by giving $K_{n}$ in terms of $A_n$ and $B_n$ instead of $K_{n-1}$ up to the first order of $\frac{\varepsilon }{2}$. By using $K_n = \frac{c_{n+1}}{c_{n}} $, even and odd terms of $c_n$ are same as (\ref{eq:22}). The power series expansion of $g(x)$ is equivalent to (\ref{eq:23}).
By using similar process from the previous case, there are two eigenvalues which are
\begin{subequations}
\begin{equation}
-\frac{\Omega }{2\mu } = n_0 -1 = |\alpha _0|  \hspace{2cm}\mbox{where}\; n_0 = 1,2,3,\cdots
\label{eq:77a}
\end{equation}
\begin{equation}
-\Big(\frac{\Omega }{2\mu } + \frac{1}{2} \Big) = n_1 -1 = |\alpha _1| \hspace{1cm} \mbox{where}\; n_1 = 1,2,3,\cdots
\label{eq:77b}
\end{equation}
\end{subequations}
(\ref{eq:77a}) makes $B_{2k+1}$ term go to zero at certain value of $k$ where $k=0,1,2,\cdots$. And (\ref{eq:77b}) makes $B_{2k+2}$ term go to zero at certain value of $k$. (\ref{eq:77a}) and (\ref{eq:77b}) make $g(x)$ function as a polynomial series. The $g(x)_{small}$ term is extremely small value relatively compared to $g(x)_{domin.}$ because it includes $A_n$ term having $\frac{\varepsilon }{2}$. First of all, add (\ref{eq:73a}) and (\ref{eq:73b}) into the first term of $g(x)$ in (\ref{eq:23}), putting $c_0 = \frac{\Gamma (\gamma +|\alpha _0|)}{\Gamma (\gamma )}$, $ \gamma = \frac{1}{2}(1+\nu ) $ and $ z=  -\frac{1}{2}\mu x^2 $.

\begin{equation}
g(x)_{ domin.} = F_{|\alpha _0|}(\gamma ;z) = \frac{\Gamma (|\alpha _{0}|+\gamma )}{\Gamma (\gamma )}\sum_{n=0}^{|\alpha _{0}|} \frac{(-|\alpha _{0}|)_n}{n!(\gamma )_n} z^n 
\label{eq:78}
\end{equation}
(\ref{eq:78}) is same as the first kind of confluent hypergeometric function. 
Substitute (\ref{eq:73a}) and (\ref{eq:73b}) into the second term of $g(x)$ in (\ref{eq:23}) using (\ref{eq:77a}) and (\ref{eq:77b}). Plug $c_0 = \frac{\Gamma (\gamma +|\alpha _0|)}{\Gamma (\gamma )}$, $ \gamma = \frac{1}{2}(1+\nu ) $ and $ z=  -\frac{1}{2}\mu x^2 $ into the new second term of $g(x)$. 
\begin{eqnarray}
g(x)_{small} &=& - x \varepsilon \sum_{n=0}^{|\alpha _0|} \frac{(n+\frac{\omega}{2})(n-\frac{1}{2})!(|\alpha _0|)!(|\alpha _0|+\gamma - \frac{1}{2})!}{(n+\gamma -\frac{1}{2})(n)!(-\frac{1}{2})!(|\alpha _0|-n)!(n+\gamma -1)!} \nonumber\\
&&\times \sum_{k=n}^{|\alpha _1|} \frac{(-1)^k (\frac{1}{2})!(n+\gamma -\frac{1}{2})!(|\alpha _1|-n)!}{(k+\frac{1}{2})!(|\alpha _1|-k)!(k+\gamma -\frac{1}{2})!} z^k
\label{eq:80}
\end{eqnarray}
As we see in (\ref{eq:80}), maximum value of index $n$ is $|\alpha _0|$. The range of index $k$ is $n\leq k\leq |\alpha _1|$. In other words, $0\leq n\leq |\alpha _0|\leq k\leq |\alpha _1|$. Then I obtain $|\alpha _0|\leq |\alpha _1|$. If $|\alpha _0|\geq  |\alpha _1|$, the function $g(x)$ will be infinite series. Then, the function $y(x)$ will blow up as I plug $g(x)$ into it no matter what the value of $x$ is. Such solution does not exist. When we see the second summation of (\ref{eq:80}), we can shift index $k$ to zero at the beginning of summation. Then we can replace the interval of index $k$ by $0\leq k\leq |\alpha _1|-n$. (\ref{eq:80}) is simply described as
\begin{eqnarray}
g(x)_{small} &=& - x \frac{\varepsilon}{2} \frac{\Gamma (|\alpha _{0}|+\gamma )}{\Gamma (\gamma )}\sum_{n=0}^{|\alpha _{0}|} \frac{(-|\alpha _{0}|)_n}{n!(\gamma )_n} z^n \nonumber\\
&&\times \sum_{k=0}^{|\alpha _{1}|-n} \frac{(n+\frac{\omega }{2})\Gamma (n+\frac{1}{2})\Gamma (n+\gamma -\frac{1}{2})(n-|\alpha _{1}|)_k}{\Gamma (k+n+\frac{3}{2})\Gamma (k+n+\gamma +\frac{1}{2})} z^k 
\label{eq:81}
\end{eqnarray}
We see the first summation of (\ref{eq:81}) which is $\frac{\Gamma (|\alpha _{0}|+\gamma )}{\Gamma (\gamma )}{\displaystyle \sum_{n=0}^{|\alpha _{0}|}} \frac{(-|\alpha _{0}|)_n}{n!\;(\gamma )_n} z^n$, it is the first kind of confluent hypergeometric polynomial of degree $|\alpha _0|$ which is denoted as $F_{|\alpha _0|}(\gamma ;z)$.
Substitute (\ref{eq:78}) and (\ref{eq:81}) into (\ref{eq:23}).
\begin{eqnarray}
g(x) &=& \mathcal{QW}_{|\alpha _{0}|,|\alpha _{1}|}\Big(|\alpha _{0}|=n_0-1,|\alpha _{1}|=n_1-1,\gamma =\frac{1}{2}(1+\nu ); z=-\frac{1}{2}\mu x^2\Big) \nonumber\\ 
&=& F_{|\alpha _0|}(\gamma ;z)- \frac{\varepsilon }{2} x \sideset{_{}^{}}{_{|\alpha _{0}|}^{|\alpha _{1}|}}\prod \big(\gamma ;z\big) \hspace{2cm} \mathrm{only}\; \mathrm{if} \; |\alpha _{0}|\leq |\alpha _{1}|
\label{eq:82}
\end{eqnarray}
where,
\begin{subequations}
\begin{equation}
F_{|\alpha _0|}(\gamma ;z) = \frac{\Gamma (|\alpha _{0}|+\gamma )}{\Gamma (\gamma )}\sum_{n=0}^{|\alpha _{0}|} \frac{(-|\alpha _{0}|)_n}{n!(\gamma )_n} z^n
\label{eq:83a}
\end{equation}
\begin{eqnarray}
\sideset{_{}^{}}{_{|\alpha _{0}|}^{|\alpha _{1}|}}\prod \big(\gamma ;z\big) &=&  \frac{\Gamma (|\alpha _{0}|+\gamma )}{\Gamma (\gamma )}\sum_{n=0}^{|\alpha _{0}|} \frac{(-|\alpha _{0}|)_n}{n!(\gamma )_n} z^n \nonumber\\
&&\times \sum_{k=0}^{|\alpha _{1}|-n} \frac{(n+\frac{\omega }{2})\Gamma (n+\frac{1}{2})\Gamma (n+\gamma -\frac{1}{2})(n-|\alpha _{1}|)_k}{\Gamma (k+n+\frac{3}{2})\Gamma (k+n+\gamma +\frac{1}{2})} z^k 
\label{eq:83b}
\end{eqnarray}
\end{subequations}
(\ref{eq:82}) denoted as
$\mathcal{QW}_{|\alpha _{0}|,|\alpha _{1}|}\Big(|\alpha _{0}|=n_0-1,|\alpha _{1}|=n_1-1,\gamma =\frac{1}{2}(1+\nu ); z=-\frac{1}{2}\mu x^2\Big)$ is the first kind of GCH polynomial of degree $|\alpha _{0}|$ and $|\alpha _{1}|$ with the first order of $\frac{\varepsilon }{2}$.
\subsubsection{Integral formalism}
The solution of the Laguerre differential equation is
\begin{equation}
L_n(z)= \sum_{k=0}^{n} \frac{(-1)^k \binom{n}{k}}{k!} z^n 
=  \frac{e^z}{n!} \frac{d^{n}}{dz^{n}} \big( z^n e^{-z}\big)
\label{eq:85}
\end{equation}
And the solution of the associated Laguerre differential equation is
\begin{equation}
L_n^k(z)= \sum_{j=0}^{n} \frac{(-1)^j (n+k)!}{(n-j)!(k+j!)(j)!} z^j = \frac{e^z z^{-k}}{n!} \frac{d^{n}}{dz^{n}} \big( z^{n+k} e^{-z}\big)
\label{eq:86}
\end{equation}
(\ref{eq:83b}) might be described in the following way:
\begin{subequations}
\begin{equation}
\sideset{_{}^{}}{_{|\alpha _{0}|}^{|\alpha _{1}|}}\prod \big(\gamma ;z\big) =  \sum_{n=0}^{|\alpha _0|} F_n^{|\alpha _0|}(\gamma ;z) \sideset{_{}^{}}{_{n}^{|\alpha _{1}|}}\prod \big(\gamma ;z\big)
\label{eq:87a}
\end{equation}
where,
\begin{eqnarray}
F_n^{|\alpha _0|}(\gamma ;z) = \frac{\Gamma (|\alpha _{0}|+\gamma )}{\Gamma (\gamma )} \frac{(-|\alpha _{0}|)_n}{n!(\gamma )_n} z^n  
\label{eq:87b}
\end{eqnarray}
\begin{eqnarray}
\sideset{_{}^{}}{_{n}^{|\alpha _{1}|}}\prod \big(\gamma ;z\big) = \sum_{k=0}^{|\alpha _{1}|-n} \frac{(n+\frac{\omega }{2})\Gamma (n+\frac{1}{2})\Gamma (n+\gamma -\frac{1}{2})(n-|\alpha _{1}|)_k}{\Gamma (k+n+\frac{3}{2})\Gamma (k+n+\gamma +\frac{1}{2})} z^k 
\label{eq:87c}
\end{eqnarray}
\end{subequations}
Plug $|\alpha _0|=0$ into (\ref{eq:82}), (\ref{eq:83a}) and (\ref{eq:87a})-(\ref{eq:87c})
\begin{equation}
\mathcal{QW}_{0,|\alpha _{1}|}\Big(\gamma ; z \Big) = 1 - \frac{\varepsilon }{2} x \sum_{k=0}^{|\alpha _{1}|} \frac{(\frac{\omega }{2})\Gamma (\frac{1}{2})\Gamma (\gamma -\frac{1}{2})(-|\alpha _{1}|)_k}{\Gamma (k+\frac{3}{2})\Gamma (k+\gamma +\frac{1}{2})} z^k 
\label{eq:88}
\end{equation}
The beta function is
\begin{equation}
B(p,q)= \frac{\Gamma (p)\Gamma (q)}{\Gamma (p+q)} = \int_0^1 dt \; t^{p-1} (1-t)^{q-1} = \int_0^{\infty } dt \frac{t^{p-1}}{(1+t)^{p+q}} = B(q,p)
\label{eq:89}
\end{equation}
And,
\begin{equation}
\frac{1}{\Gamma (k+1)} \int_{-1}^{1} dp\;(1-p^2)^k = \frac{\Gamma (\frac{1}{2})}{\Gamma (k+\frac{3}{2})}
\label{eq:90}
\end{equation}
Plug (\ref{eq:90}) into the second term of (\ref{eq:88}) on RHS
\begin{equation}
\sideset{_{}^{}}{_{0}^{|\alpha _{1}|}}\prod \big(\gamma ;z\big) = \frac{\omega }{2} \int_{-1}^1 dp \sum_{k=0}^{|\alpha _{1}|} \frac{(-1)^k \Gamma (\gamma -\frac{1}{2}) \binom{|\alpha _1|}{k}}{\Gamma (k+\gamma +\frac{1}{2})} [z(1-p^2)]^k
\label{eq:91}
\end{equation}
Replace $p$ and $q$ by $\gamma -\frac{1}{2}$ and $k+1$ in (\ref{eq:89}). Take the new (\ref{eq:89}) into (\ref{eq:91}).
\begin{equation}
\sideset{_{}^{}}{_{0}^{|\alpha _{1}|}}\prod \big(\gamma ;z\big) = \frac{\omega }{2} \int_{-1}^1 dp \int_{0}^1 dt\;t^{\gamma -\frac{3}{2}} \sum_{k=0}^{|\alpha _{1}|} \frac{(-1)^k \binom{|\alpha _1|}{k}}{\Gamma (k+1)} [z(1-t)(1-p^2)]^k
\label{eq:92}
\end{equation}
Replace $n$ and $z$ by $|\alpha _1|$ and $\varsigma = z(1-t)(1-p^2) $ in (\ref{eq:85}). Take the new (\ref{eq:85}) into (\ref{eq:92})
\begin{equation}
\sideset{_{}^{}}{_{0}^{|\alpha _{1}|}}\prod \big(\gamma ;z\big) = \frac{\omega }{2} \int_{-1}^1 dp \int_{0}^1 dt\;t^{\gamma -\frac{3}{2}} L_{|\alpha _1|}(\varsigma ) 
\label{eq:93}
\end{equation}
Plug (\ref{eq:93}) into (\ref{eq:88})
\begin{equation}
\mathcal{QW}_{0,|\alpha _{1}|}\Big(\gamma ; z \Big) =  F_{0}(\gamma ;z) - x \frac{\varepsilon }{2} \Big(\frac{\omega }{2}\Big) \int_{-1}^1 dp \int_{0}^1 dt\;t^{\gamma -\frac{3}{2}} L_{|\alpha _1|}(\varsigma ) 
\label{eq:94}
\end{equation}
Plug $|\alpha _0|=1$ into (\ref{eq:82}), (\ref{eq:83a}) and (\ref{eq:87a})-(\ref{eq:87c}) 
\begin{equation}
\mathcal{QW}_{1,|\alpha _{1}|}\Big(\gamma ; z \Big) = F_{1}(\gamma ;z)- \frac{\varepsilon }{2} x \bigg\{ F_0^1(\gamma ,z) \sideset{_{}^{}}{_{0}^{|\alpha _{1}|}}\prod \big(\gamma ;z\big) + F_1^1(\gamma ,z) \sideset{_{}^{}}{_{1}^{|\alpha _{1}|}}\prod \big(\gamma ;z\big) \bigg\}
\label{eq:95}
\end{equation}
Replace $k$ by $k+1$ in (\ref{eq:90}). Take the new (\ref{eq:90}) into the second term inside of bracket in (\ref{eq:95})
\begin{equation}
\sideset{_{}^{}}{_{1}^{|\alpha _{1}|}}\prod \big(\gamma ;z\big) = \frac{1}{2} \Big(1+\frac{\omega }{2}\Big) \int_{-1}^1 dp \;(1-p^2) \sum_{k=0}^{|\alpha _{1}|-1} \frac{(-1)^k \Gamma (|\alpha _1|)\Gamma (\gamma +\frac{1}{2})}{\Gamma (|\alpha _1|-k)\Gamma (k+\gamma +\frac{3}{2})\Gamma (k+2)} [z(1-p^2)]^k
\label{eq:96}
\end{equation}
Replace $p$ and $q$ by $\gamma + \frac{1}{2}$ and $k+1$ in (\ref{eq:89}). Take the new (\ref{eq:89}) into (\ref{eq:96}).
\begin{equation}
\sideset{_{}^{}}{_{1}^{|\alpha _{1}|}}\prod \big(\gamma ;z\big) = \frac{1}{2} \Big(1+\frac{\omega }{2}\Big) \int_{-1}^1 dp \;(1-p^2) \int_0^1 dt\; t^{\gamma -\frac{1}{2}}  \sum_{k=0}^{|\alpha _{1}|-1} \frac{(-1)^k \Gamma (|\alpha _1|)[z(1-t)(1-p^2)]^k }{\Gamma (|\alpha _1|-k)\Gamma (k+1)\Gamma (k+2)}
\label{eq:97}
\end{equation}
Replace $n, k,j$ and $z$ by $|\alpha _1|-1$, 1, $k$  and $\varsigma = z(1-t)(1-p^2) $ in (\ref{eq:86}). And take the new (\ref{eq:86}) into (\ref{eq:97}).
\begin{equation}
\sideset{_{}^{}}{_{1}^{|\alpha _{1}|}}\prod \big(\gamma ;z\big) = \frac{1}{2 |\alpha _1|}\Big(1+\frac{\omega }{2}\Big) \int_{-1}^1 dp \;(1-p^2) \int_0^1 dt\; t^{\gamma -\frac{1}{2}} L_{|\alpha _1|-1}^1(\varsigma ) 
\label{eq:98}
\end{equation}
Substitute (\ref{eq:93}) and (\ref{eq:98}) into (\ref{eq:95}).
\begin{eqnarray}
\mathcal{QW}_{1,|\alpha _{1}|}\Big(\gamma ; z \Big) &=& F_{1}(\gamma ;z)- \frac{\varepsilon }{2} x \bigg\{ F_0^1(\gamma ,z) \frac{\omega }{2} \int_{-1}^1 dp \int_{0}^1 dt\;t^{\gamma -\frac{3}{2}} L_{|\alpha _1|}(\varsigma ) \label{eq:99}\\
 &&+ F_1^1(\gamma ,z) \frac{1}{2 |\alpha _1|}\Big(1+\frac{\omega }{2}\Big) \int_{-1}^1 dp \;(1-p^2) \int_0^1 dt\; t^{\gamma -\frac{1}{2}} L_{|\alpha _1|-1}^1(\varsigma )\bigg\} 
\nonumber
\end{eqnarray}
As $|\alpha _0|=2$ in (\ref{eq:82}), (\ref{eq:83a}) and (\ref{eq:87a})-(\ref{eq:87c}), the result of its solution is similar to the previous case as $|\alpha _0|=0$ and 1.   
\begin{eqnarray}
\mathcal{QW}_{2,|\alpha _{1}|}\Big(\gamma ; z \Big) &=& F_{2}(\gamma ;z) - \frac{\varepsilon }{2} x \sum_{n=0}^2 \bigg\{ F_n^2(\gamma ,z) (n+\frac{\omega }{2}) \frac{\Gamma(n+\frac{1}{2}) \Gamma(|\alpha _1|+1-n) }{\Gamma(\frac{1}{2}) \Gamma(|\alpha _1|+1) }\nonumber\\
&\times & \int_0^1 dt\; t^{\gamma -\frac{3}{2}+n} \int_{-1}^1 dp\; (1-p^2)^n L_{|\alpha _1|-n}^n (\varsigma )\bigg\}
\label{eq:100}
\end{eqnarray}
By repeating this process, I obtain $\mathcal{QW}_{|\alpha _0|,|\alpha _{1}|}\Big(\gamma ; z \Big)$ where $|\alpha _0|\geq 3$. 
According to (\ref{eq:94}), (\ref{eq:99}) and (\ref{eq:100}), (\ref{eq:83a}) and (\ref{eq:83b}) can be described as the following way.
\begin{equation}
F_{|\alpha _0|}(z)= \frac{(|\alpha _0|)!}{2\pi i}  \oint dv \frac{e^{-\frac{zv}{(1-v)}}}{v^{|\alpha _0|+1}(1-v)^{\gamma }}
\label{eq:101}
\end{equation}
And
\begin{eqnarray}
\sideset{_{}^{}}{_{|\alpha _{0}|}^{|\alpha _{1}|}}\prod \big(\gamma ;z\big) &=& \sum_{n=0}^{|\alpha _{0}|} \bigg\{ F_n^{|\alpha _{0}|}(\gamma ,z) (n+\frac{\omega }{2}) \frac{\Gamma(n+\frac{1}{2}) \Gamma(|\alpha _1|+1-n) }{\Gamma(\frac{1}{2}) \Gamma(|\alpha _1|+1)}\nonumber\\
&&\times \int_0^1 dt\; t^{\gamma -\frac{3}{2}+n} \int_{-1}^1 dp\; (1-p^2)^n L_{|\alpha _1|-n}^n (\varsigma )\bigg\} 
\label{eq:102}
\end{eqnarray}
We know
\begin{equation}
\frac{\Gamma(n+\frac{1}{2}) \Gamma(|\alpha _1|+1-n) }{\Gamma(\frac{1}{2}) \Gamma(|\alpha _1|+1) } = \frac{B(n+\frac{1}{2}, |\alpha _1|+1-n)}{B(\frac{1}{2},|\alpha _1|+1)}
\label{eq:103}
\end{equation}
Substitute (\ref{eq:87b}) and (\ref{eq:103}) into (\ref{eq:102}).
\begin{eqnarray}
\sideset{_{}^{}}{_{|\alpha _{0}|}^{|\alpha _{1}|}}\prod \big(\gamma ;z\big) &=& \bigg\{\sum_{n=0}^{|\alpha _{0}|} \frac{(-1)^n(|\alpha _0|)!(|\alpha _0|+\gamma -1)!}{(n-1)!(|\alpha _0|-n)!(n+\gamma -1)!} \frac{B(n+\frac{1}{2}, |\alpha _1|+1-n)}{B(\frac{1}{2},|\alpha _1|+1)} \nonumber\\
&\times & \int_0^1 dt\; t^{\gamma -\frac{3}{2}} \int_{-1}^1 dp\; [z t(1-p^2)]^n L_{|\alpha _1|-n}^n (\varsigma ) \nonumber\\
&+& \frac{\omega }{2} \sum_{n=0}^{|\alpha _{0}|} \frac{(-1)^n(|\alpha _0|)!(|\alpha _0|+\gamma -1)!}{n!(|\alpha _0|-n)!(n+\gamma -1)!} \frac{B(n+\frac{1}{2}, |\alpha _1|+1-n)}{B(\frac{1}{2},|\alpha _1|+1)} \nonumber\\
&\times & \int_0^1 dt\; t^{\gamma -\frac{3}{2}}\int_{-1}^1 dp\; [z t(1-p^2)]^n L_{|\alpha _1|-n}^n (\varsigma ) \bigg\}
\label{eq:104}
\end{eqnarray}
Replace $p$ and $q$ by $n+\frac{1}{2}$ and $|\alpha _1|+1-n$ into (\ref{eq:89}). Take the new (\ref{eq:89}) into (\ref{eq:104}).
\begin{eqnarray}
\sideset{_{}^{}}{_{|\alpha _{0}|}^{|\alpha _{1}|}}\prod \big(\gamma ;z\big) &=& \frac{1}{B(\frac{1}{2},|\alpha _1|+1)}\int_0^{\infty } ds \;s^{-\frac{1}{2}}(1+s)^{-(|\alpha _1|+\frac{3}{2})}\int_0^1 dt\; t^{\gamma -\frac{3}{2}} \int_{-1}^1 dp \label{eq:105}\\
&&\times  \sum_{n=0}^{|\alpha _{0}|} \frac{(-1)^n(|\alpha _0|)!\left(n+ \frac{\omega }{2} \right)(|\alpha _0|+\gamma -1)!}{n!(|\alpha _0|-n)!(n+\gamma -1)!}  [z t s(1-p^2)]^n L_{|\alpha _1|-n}^n (\varsigma ) \nonumber
\end{eqnarray}
Integral form of Associated Laguerre polynomial is
\begin{equation}
L_n^m(z)= \frac{1}{2\pi i}  \oint du \frac{e^{-\frac{zu}{(1-u)}}}{u^{n+1}(1-u)^{m+1}}
\label{eq:106}
\end{equation}
Replace $n, m$ and $z$ by $|\alpha _1|-n$, $n$ and $\varsigma $ in (\ref{eq:106}). Take the new (\ref{eq:106}) into (\ref{eq:105}).
\begin{eqnarray}
\sideset{_{}^{}}{_{|\alpha _{0}|}^{|\alpha _{1}|}}\prod \big(\gamma ;z\big) &=& \frac{1}{2\pi i B(\frac{1}{2},|\alpha _1|+1)}\int_0^{\infty } ds \;s^{-\frac{1}{2}}(1+s)^{-(|\alpha _1|+\frac{3}{2})}\int_0^1 dt\; t^{\gamma -\frac{3}{2}} \nonumber\\
&\times &\int_{-1}^1 dp \oint du \frac{e^{-\frac{\varsigma u}{(1-u)}}}{u^{|\alpha _1|+1}(1-u)}\bigg\{zst(1-p^2)\frac{u}{1-u}\nonumber\\
&\times & \sum_{n=0}^{|\alpha _{0}|} \frac{(-1)^n(|\alpha _0|)!(|\alpha _0|+\gamma -1)!}{(n-1)!(|\alpha _0|-n)!(n+\gamma -1)!} \Big[z t s(1-p^2)\frac{u}{(1-u)}\Big]^{n-1} \nonumber\\
&+& \frac{\omega }{2} \sum_{n=0}^{|\alpha _{0}|} \frac{(-1)^n(|\alpha _0|)!(|\alpha _0|+\gamma -1)!}{n!(|\alpha _0|-n)!(n+\gamma -1)!}  \Big[z t s(1-p^2)\frac{u}{(1-u)}\Big]^n  \bigg\}
\label{eq:107}
\end{eqnarray}
Replace $z$ and $\varsigma $ by $w_1= z t s(1-p^2)\frac{u}{(1-u)}$ and $z(1-t)(1-p^2)$ into (\ref{eq:107}).
\begin{eqnarray}
\sideset{_{}^{}}{_{|\alpha _{0}|}^{|\alpha _{1}|}}\prod \big(\gamma ;z\big) &=& \frac{1}{2\pi i B(\frac{1}{2},|\alpha _1|+1)}\int_0^{\infty } ds \;s^{-\frac{1}{2}}(1+s)^{-(|\alpha _1|+\frac{3}{2})}\int_0^1 dt\; t^{\gamma -\frac{3}{2}} \nonumber\\
&\times & \int_{-1}^1 dp \oint du \frac{e^{-\frac{z u}{(1-u)}(1-t)(1-p^2)}}{u^{|\alpha _1|+1}(1-u)} \Big\{ w_1 \partial_{ w_1} + \frac{\omega }{2}\Big\} \nonumber\\
&\times & F_{|\alpha _0|}\Big(\gamma ; w_1= z t s(1-p^2)\frac{u}{(1-u)}\Big)\label{eq:108}
\end{eqnarray}
 (\ref{eq:108}) can be described as various integral forms of several different special function in the following way:
\begin{eqnarray}
\sideset{_{}^{}}{_{|\alpha _{0}|}^{|\alpha _{1}|}}\prod \big(\gamma ;z\big) &=& \frac{(|\alpha _0|)!}{2\pi i B(\frac{1}{2},|\alpha _1|+1)}\int_0^{\infty } ds \;s^{-\frac{1}{2}}(1+s)^{-(|\alpha _1|+\frac{3}{2})}\nonumber\\
 &\times &\int_0^1 dt\; t^{\gamma -\frac{3}{2}} \int_{-1}^1 dp \oint \frac{dv}{v^{|\alpha _0|+1}(1-v)^{\gamma }} \left\{ \frac{zstv(1-p^2)}{(1-v)} \partial_{ w_2} + \frac{\omega }{2}\right\} \nonumber\\
&\times &  L_{|\alpha _1|}\Big( w_2= z(1-p^2)\Big\{ (1-t)+ \frac{tsv}{(1-v)}\Big\} \Big) \label{eq:109}
\end{eqnarray}
\begin{eqnarray}
\sideset{_{}^{}}{_{|\alpha _{0}|}^{|\alpha _{1}|}}\prod \big(\gamma ;z\big) &=& \frac{(2|\alpha _0|)!}{(2\pi i)^2 B(\frac{1}{2},|\alpha _1|+1)}\int_0^{\infty } ds \;s^{-\frac{1}{2}}(1+s)^{-(|\alpha _1|+\frac{3}{2})}
\int_0^1 dt\; t^{\gamma -\frac{3}{2}} \nonumber\\
&&\times \oint \frac{dv}{v^{|\alpha _0|+1}(1-v)^{\gamma }} \oint \frac{du}{u^{|\alpha _1|+1}(1-u)} \bigg\{ \frac{zstvu}{(1-v)(1-u)} \nonumber\\
&&\times \Big( \frac{1}{3}M\big(1,\frac{5}{2},w_3\big)- M\big(1,\frac{3}{2},w_3\big) \Big) + \frac{\omega }{2}M\big(1,\frac{3}{2},w_3\big)\bigg\}  \nonumber\\
&\mbox{where},& w_3=\frac{zu}{(1-u)}\Big\{ -(1-t)-\frac{stv}{(1-v)}\Big\}
\label{eq:110}
\end{eqnarray}
\begin{eqnarray}
\sideset{_{}^{}}{_{|\alpha _{0}|}^{|\alpha _{1}|}}\prod \big(\gamma ;z\big) &=& \frac{(|\alpha _0|)!}{(2\pi i)^2 B(\frac{1}{2},|\alpha _1|+1)}\int_0^{\infty } ds \;s^{-\frac{1}{2}}(1+s)^{-(|\alpha _1|+\frac{3}{2})} \nonumber\\
&&\times \int_0^1 dp\;(1-p^2)^{-\frac{1}{2}}\oint \frac{dv}{v^{|\alpha _0|+1}(1-v)^{\gamma }} \nonumber\\
&&\times  \oint \frac{du e^{-\frac{zup}{(1-u)}}}{u^{|\alpha _1|+1}(1-u)} \bigg\{ \frac{-zspvu}{(\gamma +\frac{1}{2})(1-v)(1-u)} M\big(\gamma +\frac{1}{2},\gamma +\frac{3}{2},w_4\big) \nonumber\\
&&+ \frac{\omega }{2(\gamma -\frac{1}{2})}M\big(\gamma -\frac{1}{2},\gamma +\frac{1}{2},w_4\big)\bigg\}\nonumber\\ 
&\mbox{where},& w_4=\frac{zup}{(1-u)}\Big\{1-\frac{sv}{(1-v)}\Big\}
\label{eq:111}
\end{eqnarray}
\begin{eqnarray}
\sideset{_{}^{}}{_{|\alpha _{0}|}^{|\alpha _{1}|}}\prod \big(\gamma ;z\big) &=& \frac{\sqrt{\pi }(|\alpha _0|)!}{(2\pi i)^2 B(\frac{1}{2},|\alpha _1|+1)}\int_0^{1 } dt \;t^{\gamma -\frac{3}{2}} \int_{-1}^1 dp \nonumber\\
&\times &\oint \frac{dv}{v^{|\alpha _0|+1}(1-v)^{\gamma }}
\oint du\;\frac{ e^{-\frac{zu}{(1-u)}(1-t)(1-p^2)}}{u^{|\alpha _1|+1}(1-u)} \nonumber\\
&\times & \left\{ -\frac{ztvu(1-p^2)}{2(1-v)(1-u)} U\big(\frac{3}{2},-|\alpha _1|+1,w_5\big) + \frac{\omega }{2}U\big(\frac{1}{2},-|\alpha _1|,w_5\big)\right\} \nonumber\\ 
&\mbox{where},& w_5=\frac{ztuv}{(1-v)(1-u)}(1-p^2)
\label{eq:112}
\end{eqnarray}
In the above, $M(a,b,z)$ is the first kind of Kummer function which is
\begin{eqnarray}
M(a,b,z)&=& \sum_{n=0}^{\infty } \frac{(a)_n}{(b)_n n!} z^n = e^z M(b-a,b,-z) \nonumber\\
&=& \frac{\Gamma (b)}{\Gamma (a) \Gamma (b-a)} \int_0^1  du\;e^{zu} u^{a-1} (1-u)^{b-a-1} \nonumber\\
&\mbox{where},& \mbox{Re}(b)>\mbox{Re}(a)>0 \nonumber
\end{eqnarray}
And $U(a,b,z)$ is the second kind of Kummer function which is
\begin{eqnarray}
U(a,b,z)&=& \frac{\Gamma (1-b)}{\Gamma (a-b+1)}M(a,b,z) + \frac{\Gamma (b-1)}{\Gamma (a)} z^{1-b} M(a-b+1,2-b,z) \nonumber\\
&=& z^{1-b} U(1+a-b,2-b,z) \nonumber\\
&=& \frac{1}{\Gamma (a)}  \int_0^{\infty }  dt\;e^{-zt} t^{a-1} (1+t)^{b-a-1} \hspace{1cm} \mbox{where}, \mbox{Re}(a)>0 \nonumber
\end{eqnarray}
Substitute integral form of the first kind of confluent hypergeometric function into (\ref{eq:108}).
\begin{eqnarray}
\sideset{_{}^{}}{_{|\alpha _{0}|}^{|\alpha _{1}|}}\prod \big(\gamma ;z\big) &=& \frac{(|\alpha _0|)!}{(2\pi i)^2 B(\frac{1}{2},|\alpha _1|+1)}\int_0^{\infty } ds \;s^{-\frac{1}{2}}(1+s)^{-(|\alpha _1|+\frac{3}{2})}\nonumber\\
&\times &\int_0^1 dt\; t^{\gamma -\frac{3}{2}} \int_{-1}^1 dp \oint \frac{dv}{v^{|\alpha _0|+1}(1-v)^{\gamma }} \label{eq:115}\\ 
&\times &  \oint \frac{du}{u^{|\alpha _1|+1}(1-u)} \bigg\{-\frac{zstuv(1-p^2)}{(1-u)(1-v)}+\frac{\omega }{2} \bigg\} e^{-\frac{zu}{(1-u)}(1-p^2)\big\{ (1-t)+ \frac{tsv}{1-v}\big\}} \nonumber
\end{eqnarray}
Description I gave in (\ref{eq:109})-(\ref{eq:112}) as integral formalism is exactly equivalent (\ref{eq:115}). 
We obtain the integral representation of the first kind of GCH function by substituting (\ref{eq:101}) and (\ref{eq:115}) into (\ref{eq:82}).
\subsubsection{Generating function}
Now, let's try to get the generating function of the $1^{\mbox{st}}$ kind of GCH function. First I multiply ${\displaystyle \sum_{|\alpha _1|=|\alpha _0|}^{\infty } B\big(|\alpha _1|+1, \frac{1}{2}\big) \;v_0^{|\alpha _1|}}$ on both sides of integral form of $1^{\mbox{st}}$ kind of GCH function in (\ref{eq:82}), (\ref{eq:83a}) and (\ref{eq:83b}) where $|v_0|<1$.
\begin{equation}
{\displaystyle \sum_{|\alpha _1|=|\alpha _0|}^{\infty } B\big(|\alpha _1|+1, \frac{1}{2}\big) \;v_0^{|\alpha _1|} \mathcal{QW}_{|\alpha _{0}|,|\alpha _{1}|}\Big(\gamma =\frac{1}{2}(1+\nu ); z=-\frac{1}{2}\mu x^2\Big)= \mbox{I} -\frac{\varepsilon }{2}x \mbox{II}}
\label{eq:116}
\end{equation}
where,
\begin{subequations}
\begin{equation}
\mbox{I}= \sum_{|\alpha _1|=|\alpha _0|}^{\infty } {\displaystyle B\big(|\alpha _1|+1, \frac{1}{2}\big) v_0^{|\alpha _1|}} F_{|\alpha _0|}(z)
\label{eq:117a}
\end{equation}
\begin{equation}
\mbox{II} = \sum_{|\alpha _1|=|\alpha _0|}^{\infty } B\big(|\alpha _1|+1,  \frac{1}{2} \big) v_0^{|\alpha _1|} \sideset{_{}^{}}{_{|\alpha _{0}|}^{|\alpha _{1}|}}\prod \big(\gamma ;z\big)
\label{eq:117b}
\end{equation}
\end{subequations}
The $1^{\mbox{st}}$ kind of hypergeometric function is
\begin{eqnarray}
 _2F_1 (a,b,c,z) &=& \sum_{k=0}^{\infty } \frac{(a)_k (b)_k}{(c)_k k!} z^k = \frac{\Gamma(c) }{\Gamma(b) \Gamma(c-b) } \int_0^1 dt\;t^{b-1} (1-t)^{c-b-1}(1-zt)^{-a} \nonumber\\
 &&\mbox{only}\; \mbox{if}\; |z|<1, \; \mbox{Re}(c)>\mbox{Re}(b)>0
 \label{eq:118}
\end{eqnarray}
Substitute (\ref{eq:118}) into (\ref{eq:117a}).
\begin{eqnarray}
\mbox{I}&=& \frac{\Gamma(|\alpha _0|+1) \Gamma(\frac{1}{2}) }{\Gamma(|\alpha _0|+\frac{3}{2}) } v_0^{|\alpha _0|} \;_2 F_1 (1,|\alpha _0|+1,|\alpha _0|+\frac{3}{2},v_0) \nonumber\\
&=& \int_0^1 dt\; (v_0t)^{|\alpha _0|} (1-t)^{-\frac{1}{2}}(1-v_0t)^{-1}
\label{eq:119}
\end{eqnarray}
Substitute (\ref{eq:115}) into (\ref{eq:117b}).
\begin{eqnarray}
\mbox{II} &=& \frac{(|\alpha _0|)!}{2\pi i} \int_0^{\infty } ds \;s^{-\frac{1}{2}}(1+s)^{-\frac{3}{2}}\int_0^1 dt\; t^{\gamma -\frac{3}{2}} \int_{0}^1 dp\; (1-p)^{-\frac{1}{2}} \oint \frac{dv}{v^{|\alpha _0|+1}(1-v)^{\gamma }}\nonumber\\ 
&\times &  {\displaystyle \bigg\{\frac{1+s}{1+s-v_0}\bigg( -\frac{zvstp v_0}{(1+s-v_0)(1-v)}+\frac{\omega }{2} \bigg) \bigg\} 
e^{-\frac{zp v_0}{1+s-v_0}\Big\{ (1-t)+\frac{tsv}{1-v}\Big\}}}
\label{eq:120}
\end{eqnarray}
Multiply ${\displaystyle \sum_{|\alpha _0|=0}^{\infty } \frac{v_1^{|\alpha _0|}}{(|\alpha _0|)!}}$ with both sides of (\ref{eq:116}) where $|v_1|<1$.
\begin{eqnarray}
&& \sum_{|\alpha _0|=0}^{\infty } \sum_{|\alpha _1|=|\alpha _0|}^{\infty } \frac{B\big(|\alpha _1|+1, \frac{1}{2}\big)}{(|\alpha _0|)!} \;v_0^{|\alpha _1|} v_1^{|\alpha _0|}\mathcal{QW}_{|\alpha _{0}|,|\alpha _{1}|}\Big(\gamma =\frac{1}{2}(1+\nu ); z=-\frac{1}{2}\mu x^2\Big)\nonumber\\
&&= \mbox{A} -\frac{\varepsilon }{2}x \mbox{B}
\label{eq:121}
\end{eqnarray}
where,
\begin{subequations}
\begin{equation}
\hspace{0.6cm} \mbox{A}= \sum_{|\alpha _0|=0}^{\infty } \frac{v_1^{|\alpha _0|}}{(|\alpha _0|)!} F_{|\alpha _0|}(z) \mbox{I} 
\label{eq:122a}
\end{equation}
\begin{equation}
\mbox{B} = \sum_{|\alpha _0|=0}^{\infty } \frac{v_1^{|\alpha _0|}}{(|\alpha _0|)!} \mbox{II}
\label{eq:122b}
\end{equation}
\end{subequations}
Plug (\ref{eq:119}) into (\ref{eq:122a}).
\begin{eqnarray}
\mbox{A} &=&   \int_0^1 dt\; (1-t)^{-\frac{1}{2}}(1-v_0t)^{-1} \sum_{|\alpha _0|=0}^{\infty } \frac{(v_0 v_1t)^{|\alpha _0|}}{(|\alpha _0|)!}F_{|\alpha _0|}(z)\nonumber\\
&=&   \int_0^1 dt\; (1-t)^{-\frac{1}{2}}(1-v_0t)^{-1} (1-v_0v_1 t)^{-\gamma } {\displaystyle e^{-\frac{zv_0v_1t}{1-v_0v_1t}}} \label{eq:123}\\
&=& \sum_{n,m=0}^{\infty } \frac{(\gamma +n+m-1)! (-zv_0v_1)^n (v_0v_1)^m }{n!\;m!\;(\gamma +n-1)!} \int_0^1 dt\; t^{n+m}(1-t)^{-\frac{1}{2}}(1-v_0t)^{-1} 
\nonumber
\end{eqnarray}
Replace $a,b,c$ and $z$ by $1$, $n+m+1$, $n+m+\frac{3}{2}$ and $v_0$ into (\ref{eq:118}). And substitute the new (\ref{eq:118}) into (\ref{eq:123}).
\begin{equation}
\mbox{A} = \sum_{n,m=0}^{\infty } \frac{B\big(n+m+1,\frac{1}{2}\big) (\gamma +n)_m  (-z v_0 v_1)^n (v_0 v_1)^m}{n!\;m!} \;_2F_1 \Big(1,n+m+1,n+m+\frac{3}{2}, v_0\Big)
\label{eq:124}
\end{equation}
I can describe (\ref{eq:124}) in a different way. First, the $1^{}st$ kind of Appell hypergeometric function is given by
\begin{equation}
F_1 (\alpha ;\beta ,\beta ^{'} ;\gamma ;x,y ) = \sum_{j,k=0}^{\infty } \frac{(\alpha )_{j+k} (\beta )_j (\beta ^{'})_k}{j!\; k!\; (\gamma )_{j+k}} x^j y^k \hspace{2cm} \mbox{for}\; |x|<1, |y|<1
\label{eq:125}
\end{equation}
The function $F_1$ can be expressed by simple integral
\begin{eqnarray}
\frac{\Gamma (\alpha )\Gamma (\gamma -\alpha )}{\Gamma (\gamma)} F_1 (\alpha ;\beta ,\beta ^{'} ;\gamma ;x,y ) &=& \int_{0}^{1} du\; u^{\alpha -1} (1-u)^{\gamma -\alpha -1} (1-ux)^{-\beta } (1-uy)^{-\beta ^{'}}\nonumber\\
 && \mbox{where}\; \;\mbox{Re}(\alpha )>0, \;\mbox{Re}(\gamma -\alpha )>0
\label{eq:126} 
\end{eqnarray}
Put $\alpha = n+1$, $\gamma = n+\frac{3}{2}$, $\beta =1$, $\beta ^{'}=\gamma +n$, $x=v_0$ and $y=v_0 v_1$ in (\ref{eq:126}). And substitute the new (\ref{eq:126}) into (\ref{eq:123}).
\begin{equation}
\mbox{A} = \sum_{n=0}^{\infty } \frac{\Gamma (\frac{1}{2}) (-zv_0v_1)^n}{\Gamma (n+\frac{3}{2})} F_1 \big(n+1 ;1 ,\gamma +n ;n+\frac{3}{2} ;v_0,v_0 v_1 \big) 
\label{eq:127}
\end{equation}
(\ref{eq:127}) is exactly equivalent to the second line of (\ref{eq:124}). Substitute (\ref{eq:120}) into (\ref{eq:122b}).
\begin{equation}
\mbox{B}= \Gamma _1 (s,p,t) + \Gamma _2(s,p,t) 
\label{eq:128}
\end{equation}
where
\begin{eqnarray}
\Gamma _1(s,p,t) &=& \frac{-zv_0v_1}{(1-v_1)^{\gamma +1}} \int_0^{\infty } ds \;s^{\frac{1}{2}}(1+s)^{-\frac{1}{2}}(1+s-v_0)^{-2}\int_0^1 dt\; t^{\gamma -\frac{1}{2}}  \nonumber\\
&&\times \int_{0}^1 dp\;p (1-p)^{-\frac{1}{2}} {\displaystyle e^{-\frac{zp v_0}{1+s-v_0}\Big\{ (1-t)+\frac{tsv_1}{1-v_1}\Big\}}}
\label{eq:129}
\end{eqnarray}
\begin{eqnarray}
\Gamma _2(s,p,t) &=& \frac{\frac{\omega }{2}}{(1-v_1)^{\gamma }} \int_0^{\infty } ds \;s^{-\frac{1}{2}}(1+s)^{-\frac{1}{2}}(1+s-v_0)^{-1}\int_0^1 dt\; t^{\gamma -\frac{3}{2}} \nonumber\\
&&\times \int_{0}^1 dp\;(1-p)^{-\frac{1}{2}} {\displaystyle e^{-\frac{zp v_0}{1+s-v_0}\Big\{ (1-t)+\frac{tsv_1}{1-v_1}\Big\}}}
\label{eq:130}
\end{eqnarray}
By using the first kind of Kummer function,
\begin{eqnarray}
\int_0^1 dp\; p(1-p)^{-\frac{1}{2}} e^{-\alpha p} &=& e^{-\alpha }\bigg\{ 2M \Big( \frac{1}{2},\frac{3}{2}, \alpha \Big) -\frac{2}{3} M \Big( \frac{3}{2}, \frac{5}{2},\alpha \Big) \bigg\}\nonumber\\ & &  \mbox{where} \;\;\alpha =  \frac{z v_0}{1+s-v_0}\Big\{ (1-t)+\frac{tsv_1}{1-v_1}\Big\}
\label{eq:131}
\end{eqnarray}
Plug (\ref{eq:131}) into (\ref{eq:129}).
\begin{equation}
\Gamma _1 (s,p,t)= \gamma _1 (s,t) + \gamma _2(s,t) 
\label{eq:132}
\end{equation}
\begin{eqnarray}
\gamma _1 (s,t) &=& \frac{-2zv_0v_1}{(1-v_1)^{\gamma +1}} \int_0^{\infty } ds \;s^{\frac{1}{2}}(1+s)^{-\frac{1}{2}}(1+s-v_0)^{-2} e^{-\frac{zv_0}{1+s-v_0}} \\
&\times & {\displaystyle \int_0^1 dt\; t^{\gamma -\frac{1}{2}} e^{\frac{zv_0}{1+s-v_0}\big( 1-\frac{sv_1}{1-v_1}\big)t} M\Big( \frac{1}{2},\frac{3}{2}, \frac{z v_0}{1+s-v_0}\Big\{ (1-t)+\frac{tsv_1}{1-v_1}\Big\} \Big)} \nonumber
\label{eq:133}
\end{eqnarray}
\begin{eqnarray}
\gamma _2 (s,t) &=& \frac{2zv_0v_1}{3(1-v_1)^{\gamma +1}} \int_0^{\infty } ds \;s^{\frac{1}{2}}(1+s)^{-\frac{1}{2}}(1+s-v_0)^{-2} e^{-\frac{zv_0}{1+s-v_0}} \label{eq:134}\\
&\times & {\displaystyle \int_0^1 dt\; t^{\gamma -\frac{1}{2}} e^{\frac{zv_0}{1+s-v_0}\big( 1-\frac{sv_1}{1-v_1}\big)t} M\Big( \frac{3}{2},\frac{5}{2}, \frac{z v_0}{1+s-v_0}\Big\{ (1-t)+\frac{tsv_1}{1-v_1}\Big\} \Big)} \nonumber
\end{eqnarray}
There are addition theorem for $M(a,b,x+y)$.
\begin{subequations}
\begin{equation}
M(a,b,x+y)= \sum_{n=0}^{\infty } \frac{(a)_n y^n}{(b)_n n!} M(a+n,b+n,x)
\label{eq:135a}
\end{equation}
\begin{equation}
\hspace{1cm}M(a,b,x+y)= e^y \sum_{n=0}^{\infty } \frac{(b-a)_n (-y)^n}{(b)_n n!} M(a,b+n,x)
\label{eq:135b}
\end{equation}
\end{subequations}
Put $a=\frac{3}{2}$, $b=\frac{5}{2}$, $x=\frac{zv_0}{1+s-v_0}$ and $y= -\frac{zv_0}{1+s-v_0} \Big( 1-\frac{sv_1}{1-v_1}\Big)t$ into (\ref{eq:135b}), and take the new (\ref{eq:135b}) into (\ref{eq:133}) and replace $s$ by $\frac{1}{t}-1$ in it.
 \begin{eqnarray}
\gamma _1 (s,t) &=& 2 \sum_{n=0}^{\infty } \frac{(1-v_1)^{-(n+\gamma +1)}(zv_0)^{n+1}(-v_1)^{n+1}}{(\gamma +n+\frac{1}{2})(\frac{3}{2})_n} 
 \label{eq:136}\\
&&\times \int_0^{1 } dt \;(1-t)^{\frac{1}{2}}(1-v_0t)^{-(n+2)}\Big(1-\frac{1}{v_1}t\Big)^n {\displaystyle M\Big( n+1,n+\frac{3}{2}, -\frac{z t v_0}{1-v_0t}\Big)} 
\nonumber
\end{eqnarray}
Put $a=\frac{3}{2}$, $b=\frac{5}{2}$, $x=\frac{zv_0}{1+s-v_0}$ and $y= -\frac{zv_0}{1+s-v_0} \Big( 1-\frac{sv_1}{1-v_1}\Big)t$ into (\ref{eq:135b}), and take the new (\ref{eq:135b}) into (\ref{eq:134}) and replace $s$ by $\frac{1}{t}-1$ in it.
\begin{eqnarray}
\gamma _2 (s,t) &=& -\frac{2}{3} \sum_{n=0}^{\infty } \frac{(1-v_1)^{-(n+\gamma +1)}(zv_0)^{n+1}(-v_1)^{n+1}}{(\gamma +n+\frac{1}{2})(\frac{5}{2})_n} \label{eq:137}\\
&&\times \int_0^{1} dt \;(1-t)^{\frac{1}{2}}(1-v_0t)^{-(n+2)}\Big(1-\frac{1}{v_1}t\Big)^n {\displaystyle M\Big( n+1,n+\frac{5}{2}, -\frac{z t v_0}{1-v_0t}\Big)} 
\nonumber
\end{eqnarray}
We know that
\begin{subequations}
\begin{equation}
M\Big(n+1,n+\frac{3}{2},\frac{-zv_0t}{1-v_0t} \Big)= \sum_{m=0}^{\infty } \frac{(n+1)_m (-zv_0)^m}{(n+\frac{3}{2})_m\;m!} t^m (1-v_0t)^{-m}
\label{eq:138a}
\end{equation}
\begin{equation}
M\Big(n+1,n+\frac{5}{2},\frac{-zv_0t}{1-v_0t} \Big)= \sum_{m=0}^{\infty } \frac{(n+1)_m (-zv_0)^m}{(n+\frac{5}{2})_m\;m!} t^m (1-v_0t)^{-m}
\label{eq:138b}
\end{equation}
\end{subequations}
Plug (\ref{eq:138a}) and (\ref{eq:138b}) into(\ref{eq:136}) and (\ref{eq:137}). Put $\alpha = m+1$, $\beta =n+m+2$, $\beta ^{'}=-n$, $\gamma =m+\frac{5}{2}$, $x=v_0$ and $y=\frac{1}{v_1}$ in (\ref{eq:126}). And substitute the new (\ref{eq:126}) into the new (\ref{eq:136}) and (\ref{eq:137}).
\begin{eqnarray}
\gamma _1 (s,t) &=& 2\sum_{n,m=0}^{\infty } \frac{(-1)^{n+m+1}(1-v_1)^{-(n+\gamma +1)}(zv_0)^{n+m+1}(v_1)^{n+1}\Gamma (\frac{3}{2})}{(\gamma +n+\frac{1}{2})(\frac{3}{2})_n (n+\frac{3}{2})_m \Gamma (m+\frac{5}{2})}\nonumber\\
&&\times  F_1 \Big(m+1; n+m+2, -n;m+\frac{5}{2}; v_0,\frac{1}{v_1}\Big)
\label{eq:139}
\end{eqnarray}
\begin{eqnarray}
\gamma _2 (s,t) &=& -\frac{2}{3}\sum_{n,m=0}^{\infty } \frac{(-1)^{n+m+1}(1-v_1)^{-(n+\gamma +1)}(zv_0)^{n+m+1}(v_1)^{n+1}\Gamma (\frac{3}{2})}{(\gamma +n+\frac{1}{2})(\frac{5}{2})_n (n+\frac{5}{2})_m \Gamma (m+\frac{5}{2})} \nonumber\\
&&\times F_1 \Big(m+1; n+m+2, -n;m+\frac{5}{2}; v_0,\frac{1}{v_1}\Big)
\label{eq:140}
\end{eqnarray}
Substitute (\ref{eq:139}) and (\ref{eq:140}) into (\ref{eq:132}).
\begin{eqnarray}
\Gamma _1 (s,p,t) &=& \sum_{n,m=0}^{\infty } \frac{(-1)^{n+m+1}(n+m+1)\Gamma (\frac{1}{2})\Gamma(\frac{3}{2}) (1-v_1)^{-(n+\gamma +1)}(zv_0)^{n+m+1}(v_1)^{n+1}}{(\gamma +n+\frac{1}{2})\Gamma (m+\frac{5}{2}) \Gamma (n+m+\frac{5}{2}) }\nonumber\\
&&\times   F_1 \Big(m+1; n+m+2, -n;m+\frac{5}{2}; v_0,\frac{1}{v_1}\Big)
\label{eq:141}
\end{eqnarray}
Replace $s$ and $p$ by $\frac{1}{v}-1$ and $1-p^2$ in (\ref{eq:130}).
\begin{eqnarray}
\Gamma _2(s,p,t) &=& \frac{\frac{\omega }{2}}{(1-v_1)^{\gamma }} \int_0^{1} dv \;(1-v)^{-\frac{1}{2}}(1-v_0 v)^{-1}\int_0^1 dt\; t^{\gamma -\frac{3}{2}}{\displaystyle e^{-\frac{z v_0 v}{1-v_0 v}\Big\{ (1-t)+\frac{tv_1 (1-v)}{v(1-v_1)}\Big\}}}\nonumber\\
&&\times  \int_{-1}^1 dp\;{\displaystyle e^{\frac{z v_0 v}{1-v_0 v}\Big\{ (1-t)+\frac{tv_1 (1-v)}{v(1-v_1)}\Big\}p^2}}\nonumber\\
&=& \frac{\omega }{(1-v_1)^{\gamma }} \int_0^{1} dv \;(1-v)^{-\frac{1}{2}}(1-v_0 v)^{-1}\int_0^1 dt\; t^{\gamma -\frac{3}{2}}\nonumber\\
&&\times  M\Big( 1,\frac{3}{2}, \frac{-zv_0 v}{1-v_0 v}\Big\{ 1-t+ \frac{v_1(1-v)t}{v(1-v_1)} \Big\} \Big)
\label{eq:142}
\end{eqnarray}
Put $a=1$, $b=\frac{3}{2}$, $x=-\frac{zv_0}{1-v_0 v}$ and $y= \frac{zv_0 v}{1-v_0 v}\Big\{ 1- \frac{v_1(1-v)}{v(1-v_1)} \Big\}t$ into (\ref{eq:135a}), and take the new (\ref{eq:135a}) into (\ref{eq:142}).
 \begin{eqnarray}
\Gamma _2 (s,p,t) &=& \frac{\omega }{(1-v_1)^{\gamma }} \sum_{n,m=0}^{\infty } \frac{v_1^n (1-v_1)^{-n}(-zv_0)^{n+m}(n+1)_m}{(\gamma +n-\frac{1}{2})(\frac{3}{2})_n (n+\frac{3}{2})_m\; m! }\nonumber\\
 &&\times \int_0^1 dv\;v^m (1-v)^{-\frac{1}{2}} (1-v_0 v)^{-(n+m+1)} \Big(1-\frac{v}{v_1}\Big)^n
 \label{eq:143}
\end{eqnarray}
Put $\alpha = m+1$, $\gamma  =m+\frac{3}{2}$, $\beta ^{'}=-n$, $\beta  =n+m+1$, $x=v_0$ and $y=\frac{1}{v_1}$ in (\ref{eq:126}). Take the new (\ref{eq:126}) into (\ref{eq:143}). 
\begin{eqnarray}
\Gamma _2 (s,p,t) &=& \omega \sum_{n,m=0}^{\infty } \frac{(-1)^{n+m}\Gamma (\frac{1}{2})(n+1)_m(1-v_1)^{-(n+\gamma )}(zv_0)^{n+m}(v_1)^{n}}{(\gamma +n-\frac{1}{2})(\frac{3}{2})_n (n+\frac{3}{2})_m \Gamma (m+\frac{3}{2})}\nonumber\\
&&\times  F_1 \Big(m+1; n+m+1, -n;m+\frac{3}{2}; v_0,\frac{1}{v_1}\Big)
 \label{eq:144}
\end{eqnarray}
Substitute (\ref{eq:141}) and (\ref{eq:144}) into (\ref{eq:128}).
\begin{eqnarray}
\mbox{B}&=& \sum_{n,m=0}^{\infty } \frac{(-1)^{n+m+1}(n+m+1)\Gamma (\frac{1}{2})\Gamma(\frac{3}{2}) (1-v_1)^{-(n+\gamma +1)}(zv_0)^{n+m+1}(v_1)^{n+1}}{(\gamma +n+\frac{1}{2})\Gamma (m+\frac{5}{2}) \Gamma (n+m+\frac{5}{2}) }\nonumber\\
&&\times F_1 \Big(m+1; n+m+2, -n;m+\frac{5}{2}; v_0,\frac{1}{v_1}\Big) \nonumber\\
&&+ \omega \sum_{n,m=0}^{\infty } \frac{(-1)^{n+m}\Gamma (\frac{1}{2})(n+1)_m(1-v_1)^{-(n+\gamma )}(zv_0)^{n+m}(v_1)^{n}}{(\gamma +n-\frac{1}{2})(\frac{3}{2})_n (n+\frac{3}{2})_m \Gamma (m+\frac{3}{2})} \nonumber\\
&&\times F_1 \Big(m+1; n+m+1, -n;m+\frac{3}{2}; v_0,\frac{1}{v_1}\Big)
\label{eq:145}
\end{eqnarray}
Plug (\ref{eq:127}) and (\ref{eq:145}) into (\ref{eq:121}). Then I obtain the generating function for $1^{\mbox{st}}$ kind of GCH function in the following way.
\begin{eqnarray}
&& \sum_{|\alpha _0|=0}^{\infty } \sum_{|\alpha _1|=|\alpha _0|}^{\infty } \frac{B\big(|\alpha _1|+1, \frac{1}{2}\big)}{(|\alpha _0|)!} \;v_0^{|\alpha _1|} v_1^{|\alpha _0|}\mathcal{QW}_{|\alpha _{0}|,|\alpha _{1}|}\Big(\gamma =\frac{1}{2}(1+\nu ); z=-\frac{1}{2}\mu x^2\Big)\nonumber\\
&=&  \sum_{n=0}^{\infty } \frac{\Gamma (\frac{1}{2}) (-zv_0v_1)^n}{\Gamma (n+\frac{3}{2})} F_1 \big(n+1 ;1 ,\gamma +n ;n+\frac{3}{2} ;v_0,v_0 v_1 \big) 
\nonumber\\
&&- \frac{\varepsilon}{2} x \Bigg\{ \sum_{n,m=0}^{\infty } \frac{(-1)^{n+m+1}(n+m+1)\Gamma (\frac{1}{2})\Gamma(\frac{3}{2}) (1-v_1)^{-(n+\gamma +1)}(zv_0)^{n+m+1}(v_1)^{n+1}}{(\gamma +n+\frac{1}{2})\Gamma (m+\frac{5}{2}) \Gamma (n+m+\frac{5}{2}) }\nonumber\\
&&\times F_1 \Big(m+1; n+m+2, -n;m+\frac{5}{2}; v_0,\frac{1}{v_1}\Big)\nonumber\\
&&+ \omega \sum_{n,m=0}^{\infty } \frac{(-1)^{n+m}\Gamma (\frac{1}{2})(n+1)_m(1-v_1)^{-(n+\gamma )}(zv_0)^{n+m}(v_1)^{n}}{(\gamma +n-\frac{1}{2})(\frac{3}{2})_n (n+\frac{3}{2})_m \Gamma (m+\frac{3}{2})}\nonumber\\
&&\times F_1 \Big(m+1; n+m+1, -n;m+\frac{3}{2}; v_0,\frac{1}{v_1}\Big) \Bigg\}
\label{eq:146}
\end{eqnarray}
Also, I can describe the generating function of it as the integral formalism by using (\ref{eq:123}), (\ref{eq:129}) and (\ref{eq:130}). 

\begin{eqnarray}
 && \sum_{|\alpha _0|=0}^{\infty } \sum_{|\alpha _1|=|\alpha _0|}^{\infty } \frac{B\big(|\alpha _1|+1, \frac{1}{2}\big)}{(|\alpha _0|)!} \;v_0^{|\alpha _1|} v_1^{|\alpha _0|}\mathcal{QW}_{|\alpha _{0}|,|\alpha _{1}|}\Big(\gamma =\frac{1}{2}(1+\nu ); z=-\frac{1}{2}\mu x^2\Big)\nonumber\\
&=&  \int_0^1 dt\; (1-t)^{-\frac{1}{2}}(1-v_0t)^{-1} (1-v_0v_1 t)^{-\gamma } {\displaystyle e^{-\frac{zv_0v_1t}{1-v_0v_1t}}} \nonumber\\
&&-\frac{\varepsilon }{2} \frac{x}{(1-v_1)^{\gamma }}\int_0^{1} du \;(1-u)^{-\frac{1}{2}}(1-v_0u)^{-1}\int_0^1 dt\; t^{\gamma -\frac{3}{2}} \int_{0}^1 dp\;(1-p)^{-\frac{1}{2}}\nonumber\\
&&\times \bigg\{ \frac{\omega }{2}- \frac{zpt(1-u)v_0 v_1}{(1-v_1)(1-u v_1)}\bigg\} e^{-\frac{zpu v_0}{1-u v_0}\Big\{1-\frac{t(u-v_1)}{u(1-v_1)}\Big\}}
 \label{eq:147}
\end{eqnarray}
\subsubsection{Orthogonal relation}
From the differential equations satisfied by $\mathcal{QW}_{|\alpha _{0}|,|\alpha _{1}|}(z)$ and $\mathcal{QW}_{|\beta  _{0}|,|\beta  _{1}|}(z)$, namely,
\begin{subequations}
\begin{equation}
x \mathcal{QW}_{|\alpha _{0}|,|\alpha _{1}|}^{''}(z) + (\mu x^2 + \varepsilon x + \nu ) \mathcal{QW}_{|\alpha _{0}|,|\alpha _{1}|}^{'}(z) + (\Omega _{|\alpha _0|,|\alpha _1|} x + \varepsilon \omega  )\mathcal{QW}_{|\alpha _{0}|,|\alpha _{1}|}(z) = 0
\label{eq:148a}
\end{equation}
\begin{equation}
x \mathcal{QW}_{|\beta _{0}|,|\beta _{1}|}^{''}(z) + (\mu x^2 + \varepsilon x + \nu ) \mathcal{QW}_{|\beta _{0}|,|\beta  _{1}|}^{'}(z) + (\Omega _{|\beta _0|,|\beta  _1|} x + \varepsilon \omega  )\mathcal{QW}_{|\beta  _{0}|,|\beta  _{1}|}(z) = 0
\label{eq:148b}
\end{equation}
\end{subequations}
where,
\begin{equation}
\begin{array}{ll}
\Omega _{|\alpha _0|,|\alpha _1|} =  \left\{\begin{array}{c}
-2\mu |\alpha _0|\hspace{1cm} \\
-2\mu (|\alpha  _1|+\frac{1}{2})
\end{array} \right.
\hspace{2cm}\Omega _{|\beta _0|,|\beta  _1|} = 
\left\{\begin{array}{c}
-2\mu |\beta _0|\hspace{1cm}\\
-2\mu (|\beta _1|+\frac{1}{2})
\end{array}\right.
\end{array}
\label{eq:149}
\end{equation}
multiplying (\ref{eq:148a}) and (\ref{eq:148b}) by $\mathcal{QW}_{|\beta  _{0}|,|\beta  _{1}|}(z)$ and $\mathcal{QW}_{|\alpha _{0}|,|\alpha _{1}|}(z)$ respectively and subtracting, I have
\begin{equation}
x \frac{d M(z)}{dx}+ (\mu x^2 + \varepsilon x + \nu )M(z) = (\Omega _{|\beta  _0|,|\beta  _1|}- \Omega _{|\alpha _0|,|\alpha _1|})x \mathcal{QW}_{|\alpha _{0}|,|\alpha _{1}|}(z) \mathcal{QW}_{|\beta  _{0}|,|\beta  _{1}|}(z)
\label{eq:150}
\end{equation}
where
\begin{equation}
M(z)= \mathcal{QW}_{|\alpha _{0}|,|\alpha _{1}|}^{'}(z) \mathcal{QW}_{|\beta  _{0}|,|\beta  _{1}|}(z) - \mathcal{QW}_{|\beta _{0}|,|\beta  _{1}|}^{'}(z)\mathcal{QW}_{|\alpha _{0}|,|\alpha _{1}|}(z) 
\label{eq:151}
\end{equation}
Multiply $x^{\nu -1 }e^{\frac{1}{2}\mu x^2+\varepsilon x}$ on both side of (\ref{eq:150}) Then integrate it with respect to $x$ from 0 to $\infty $.
\begin{eqnarray}
 &&(\Omega _{|\beta  _0|,|\beta  _1|}- \Omega _{|\alpha _0|,|\alpha _1|})\int_0^{\infty }dx\;x^{\nu }e^{\frac{1}{2}\mu x^2+\varepsilon x} \mathcal{QW}_{|\alpha _{0}|,|\alpha _{1}|}(z) \mathcal{QW}_{|\beta  _{0}|,|\beta  _{1}|}(z)  \label{eq:152}\\
 &=&\Big[x^{\nu }e^{\frac{1}{2}\mu x^2+\varepsilon x} \big( \mathcal{QW}_{|\alpha _{0}|,|\alpha _{1}|}^{'}(z) \mathcal{QW}_{|\beta  _{0}|,|\beta  _{1}|}(z) - \mathcal{QW}_{|\beta _{0}|,|\beta  _{1}|}^{'}(z)\mathcal{QW}_{|\alpha _{0}|,|\alpha _{1}|}(z)\big)\Big]\bigg|_0^{\infty }
\nonumber
\end{eqnarray}
Therefore, only if $|\alpha _0| \neq |\beta _0|$ and $|\alpha _1| \neq |\beta _1|$, then
\begin{equation}
\int_0^{\infty }dx\;x^{\nu }e^{\frac{1}{2}\mu x^2+\varepsilon x} \mathcal{QW}_{|\alpha _{0}|,|\alpha _{1}|}(z) \mathcal{QW}_{|\beta  _{0}|,|\beta  _{1}|}(z) = 0 
\label{eq:153}
\end{equation}
Let's think about the case in which are $|\alpha _0| = |\beta _0|$ and $|\alpha _1| = |\beta _1|$. First of all, from (\ref{eq:82})
\begin{eqnarray}
\Big[\mathcal{QW}_{|\alpha _{0}|,|\alpha _{1}|}(z)\Big]^2 &=& \Big[ F_{|\alpha _0|}(z)\Big]^2 -\varepsilon x F_{|\alpha _0|}(z) \sideset{_{}^{}}{_{|\alpha _{0}|}^{|\alpha _{1}|}}\prod \big(z\big) + \frac{\varepsilon ^2}{4} x^2 \Big[ \sideset{_{}^{}}{_{|\alpha _{0}|}^{|\alpha _{1}|}}\prod \big(z\big)\Big]^2 \nonumber\\
&\simeq & \Big[ F_{|\alpha _0|}(z)\Big]^2 -\varepsilon x F_{|\alpha _0|}(z) \sideset{_{}^{}}{_{|\alpha _{0}|}^{|\alpha _{1}|}}\prod \big(z\big)
\label{eq:154}
\end{eqnarray}
$\varepsilon $ is extremely small. So anything beyond the second order of $\varepsilon $ is negligible in (\ref{eq:154}). Multiply $x^{\nu }e^{\frac{1}{2}\mu x^2+\varepsilon x}$ in (\ref{eq:154}) and integrate them with respect to $x=[0,\infty ]$.
\begin{equation}
\int_0^{\infty }dx\;x^{\nu }e^{\frac{1}{2}\mu x^2+\varepsilon x} \Big[\mathcal{QW}_{|\alpha _{0}|,|\alpha _{1}|}(z)\Big]^2 \approx E_1 -\varepsilon E_2 
\label{eq:155}
\end{equation}
where
\begin{subequations}
\begin{equation}
E_1= \int_0^{\infty }dx\;x^{\nu }e^{\frac{1}{2}\mu x^2+\varepsilon x} \Big[ F_{|\alpha _0|}(z)\Big]^2 \hspace{1.5cm}
\label{eq:156a}
\end{equation}
\begin{equation}
E_2=\int_0^{\infty }dx\;x^{\nu+1 }e^{\frac{1}{2}\mu x^2+\varepsilon x} F_{|\alpha _0|}(z) \sideset{_{}^{}}{_{|\alpha _{0}|}^{|\alpha _{1}|}}\prod \big(z\big)
\label{eq:156b}
\end{equation}
\end{subequations}
Apply the generating function of the first kind of confluent hypergeometric function into (\ref{eq:156a}) and (\ref{eq:156b}). 
\begin{equation}
E_1 \approx  \frac{2^{\gamma -1}}{|\mu |^{\gamma }}\Gamma (|\alpha _0|+1) \Gamma (|\alpha _0|+\gamma ) + \varepsilon \frac{(-1)^{|\alpha _0|}2^{\gamma -\frac{1}{2}} \Gamma \Big( |\alpha _0|+\gamma -\frac{1}{2}\Big) \Gamma \Big( |\alpha _0|+\gamma +\frac{1}{2}\Big)}{|\mu |^{\gamma +\frac{1}{2}}\Gamma \Big( \gamma -\frac{1}{2}\Big)}
\label{eq:157}
\end{equation}
\begin{eqnarray}
E_2 &\approx & \frac{2^{\gamma -\frac{3}{2}}}{|\mu |^{\gamma +\frac{1}{2}}} \frac{\Gamma (|\alpha _{0}|+\gamma )}{\Gamma (\gamma )}\sum_{n=0}^{|\alpha _{0}|} \frac{(-|\alpha _{0}|)_n}{n!(\gamma )_n} \sum_{k=0}^{|\alpha _{1}|-n} \frac{(n+\frac{\omega }{2})\Gamma (n+\frac{1}{2})\Gamma (n+\gamma -\frac{1}{2})(n-|\alpha _{1}|)_k}{\Gamma (k+n+\frac{3}{2})\Gamma (k+n+\gamma +\frac{1}{2})} \nonumber\\ 
&&\times \Bigg\{ \frac{(-1)^{|\alpha _0|}\Gamma \big(n+k+\frac{3}{2}\big)\Gamma \big( n+k+\gamma +\frac{1}{2}\big)}{\Gamma \big( n+k-|\alpha _0|+\frac{3}{2}\big)}\nonumber\\
&&+ \varepsilon \frac{2}{\sqrt{2|\mu |}}\frac{(-1)^{|\alpha _0|}\Gamma (n+k+\gamma +1) \Gamma (n+k+2)}{\Gamma (n+k-|\alpha _0|+2)}\Bigg\}
\label{eq:158}
\end{eqnarray}
I neglect any terms which include more than second order of $\varepsilon $, extremely small, in (\ref{eq:157}) and (\ref{eq:158}).
Substitute (\ref{eq:157}) and (\ref{eq:158}) into (\ref{eq:155}) with neglecting any terms larger than second order of $\varepsilon $.
Therefore,the orthogonal relation of it is
\begin{eqnarray}
 && \int_0^{\infty }dx\;x^{\nu }e^{\frac{1}{2}\mu x^2+\varepsilon x} \mathcal{QW}_{|\alpha _{0}|,|\alpha _{1}|}(z) \mathcal{QW}_{|\beta  _{0}|,|\beta _{1}|}(z)\nonumber\\ 
&\approx & \Bigg\{ \frac{2^{\gamma -1}}{|\mu |^{\gamma }}\Gamma (|\alpha _0|+1) \Gamma (|\alpha _0|+\gamma )
 + \varepsilon \frac{(-1)^{|\alpha _0|}2^{\gamma -\frac{1}{2}} }{|\mu |^{\gamma +\frac{1}{2}}} \Bigg\{ \frac{\Gamma \Big( |\alpha _0|+\gamma -\frac{1}{2}\Big) \Gamma \Big( |\alpha _0|+\gamma +\frac{1}{2}\Big)}{ \Gamma \Big( \gamma -\frac{1}{2}\Big)} \nonumber\\
&&-  \frac{\Gamma (|\alpha _{0}|+\gamma )}{\Gamma (\gamma )}\sum_{n=0}^{|\alpha _{0}|} \frac{(-|\alpha _{0}|)_n}{n!(\gamma )_n}\label{eq:160}\\
&&\times \sum_{k=0}^{|\alpha _{1}|-n} \frac{(n+\frac{\omega }{2})\Gamma (n+\frac{1}{2})\Gamma (n+\gamma -\frac{1}{2})(n-|\alpha _{1}|)_k}{2 \Gamma(k+n+\frac{3}{2}-|\alpha_0|)} \Bigg\}\Bigg\} \delta _{|\alpha _0|,|\beta _0|} \;\delta _{|\alpha _1|,|\beta _1|}\nonumber
\end{eqnarray}

Also, if there is an analytic function $\Psi (x)$ having first and second continuous derivatives in $[0,\infty ]$ and approaching zero when $x\rightarrow \infty $, they can be expanded in terms of $\mathcal{QW}_{|\alpha _0|,|\alpha _1|}( \gamma ; z)$ by using (\ref{eq:160}):
\begin{equation}
\Psi (x) = \sum_{|\alpha _0|=0}^{\infty } \sum_{|\alpha _1|=|\alpha _0|}^{\infty } A_{|\alpha _0|,|\alpha _1|} \mathcal{QW}_{|\alpha _0|,|\alpha _1|}\Big( \gamma =\frac{1}{2}(1+\nu ); z=-\frac{1}{2}\mu x^2\Big) 
\label{eq:161}
\end{equation}
where,
\begin{eqnarray}
A_{|\alpha _0|,|\alpha _1|} &\approx & \int_{0}^{\infty } dx\;x^{\nu } e^{\frac{1}{2}\mu x^2+ \varepsilon x}\mathcal{QW}_{|\alpha _0|,|\alpha _1|}( \gamma ; z) \Psi (x) \Bigg[ \frac{2^{\gamma -1}}{|\mu |^{\gamma }}\Gamma (|\alpha _0|+1) \Gamma (|\alpha _0|+\gamma )\nonumber\\
&&+ \varepsilon \frac{(-1)^{|\alpha _0|}2^{\gamma -\frac{1}{2}} }{|\mu |^{\gamma +\frac{1}{2}}}\Bigg\{ \frac{\Gamma ( |\alpha _0|+\gamma -\frac{1}{2}) \Gamma ( |\alpha _0|+\gamma +\frac{1}{2})}{ \Gamma ( \gamma -\frac{1}{2})}\label{eq:162}\\
&&- \frac{\Gamma (|\alpha _{0}|+\gamma )}{\Gamma (\gamma )}\sum_{n=0}^{|\alpha _{0}|} \frac{(-|\alpha _{0}|)_n}{n!(\gamma )_n} \sum_{k=0}^{|\alpha _{1}|-n} \frac{(n+\frac{\omega }{2})\Gamma (n+\frac{1}{2})\Gamma (n+\gamma -\frac{1}{2})(n-|\alpha _{1}|)_k}{2 \Gamma(k+n+\frac{3}{2}-|\alpha _0|)} \Bigg\} \Bigg]^{-1}
\nonumber
\end{eqnarray}
\subsection{ As $\lambda _2 = 1-\nu $ }
\subsubsection{Power series expansion in closed forms and integral formalism}
In the previous case, I obtain the first kind of independent solution of GCH function, putting $\lambda _1 = 0 $.

Now put $\lambda _2 = 1-\nu $ into (\ref{eq:43}) to get the second solution of GCH function.
\begin{subequations}
\begin{equation}
A_n|_{\lambda =1-\nu}= - \frac{\varepsilon (n+\omega -\nu +1 )}{(n+1)(n+2-\nu )}
\label{eq:163a}
\end{equation}
\begin{equation}
B_n|_{\lambda =1-\nu}= - \frac{\Omega +\mu (n-\nu )}{(n+1)(n+2-\nu )}
\label{eq:163b}
\end{equation}
\end{subequations}
And there are eigenvalues where $ \gamma =\frac{1}{2}(1+\nu )$ which are
\begin{subequations}
\begin{equation}
-\frac{\Omega }{2\mu } +\gamma -1 = \psi_0   \hspace{2cm}\mbox{where}\;  \psi _0  = 1,2,3,\cdots
\label{eq:164a}
\end{equation}
\begin{equation}
-\frac{\Omega }{2\mu }+ \gamma -\frac{3}{2} = \psi _1 \hspace{2cm} \mbox{where}\;  \psi _1 = 1,2,3,\cdots
\label{eq:164b}
\end{equation}
\end{subequations}
Substitute (\ref{eq:163a}) and (\ref{eq:163b}) into (\ref{eq:23}). Then I obtain the second independent solution of GCH function by using similar process as I did before. 
\begin{equation}
\mathcal{RW}_{\psi _0,\psi _1}\Big( \gamma =\frac{1}{2}(1+\nu ); z=-\frac{1}{2}\mu x^2\Big) = z^{1-\gamma }\Bigg\{ A_{\psi _0}(\gamma ;z)- \frac{\varepsilon }{2} x  \sideset{_{}^{}}{_{\psi _0}^{\psi _1}}\bigwedge \big(\gamma ;z\big)\Bigg\}
 \label{eq:166}
\end{equation}
where, 
\begin{equation}
A_{\psi _0}(\gamma ;z) = \frac{\Gamma (\psi _0+2-\gamma )}{\Gamma (2-\gamma )}\sum_{n=0}^{\psi _0} \frac{(-\psi _0)_n}{n!(2-\gamma )_n} z^{n}
\label{eq:167}
\end{equation}
\begin{eqnarray}
 \sideset{_{}^{}}{_{\psi _0}^{\psi _1}}\bigwedge \big(\gamma ;z\big) &=&  \frac{\Gamma (\psi _0+2-\gamma )}{\Gamma (2-\gamma )}\sum_{n=0}^{\psi _0} \frac{(-\psi _0)_n}{n!(2-\gamma )_n} z^{n} \label{eq:300} \\ 
&&\times  \sum_{k=0}^{\psi _1 -n} \frac{(\frac{\omega }{2}+n+1-\gamma )\Gamma (n+\frac{1}{2})\Gamma (n+\frac{3}{2}-\gamma)(n-\psi _1)_k}{\Gamma (k+n+\frac{3}{2})\Gamma (k+n+\frac{5}{2}-\gamma )} z^{k} \nonumber
\end{eqnarray}
(\ref{eq:300}) can be described as various integral forms of several different special function in the following way:
\begin{eqnarray}
\sideset{_{}^{}}{_{\psi _0}^{\psi _1}}\bigwedge \big(\gamma ;z\big) &=&  \frac{1}{2\pi i B(\frac{1}{2},\psi _1+1)}\int_0^{\infty } ds \;s^{-\frac{1}{2}}(1+s)^{-(\psi _1+\frac{3}{2})}\nonumber\\
&&\times \int_0^1 dt\; t^{\frac{1}{2}-\gamma } \int_{-1}^1 dp \oint du \frac{e^{-\frac{zu}{(1-u)}(1-t)(1-p^2)}}{u^{\psi _1+1}(1-u)} \label{eq:168a}\\
&&\times \Big\{ w_1 \partial_{w_1} + \Big( \frac{\omega }{2}+1-\gamma  \Big) \Big\} A_{\psi _0}\Big(\gamma ; w_1= z t s(1-p^2)\frac{u}{(1-u)}\Big)
\nonumber
\end{eqnarray}
\begin{eqnarray}
\sideset{_{}^{}}{_{\psi _0}^{\psi _1}}\bigwedge \big(\gamma ;z\big) &=&  \frac{\psi _0!}{2\pi i B(\frac{1}{2},\psi _1 +1)}\int_0^{\infty } ds \;s^{-\frac{1}{2}}(1+s)^{-(\psi _1+\frac{3}{2})}\int_0^1 dt\; t^{\frac{1}{2}-\gamma } \nonumber\\
&& \times \int_{-1}^1 dp \oint \frac{dv}{v^{\psi _0+1}(1-v)^{2-\gamma }} \Big\{ \frac{zstv(1-p^2)}{(1-v)} \partial_{ w_2} + \Big( \frac{\omega }{2}+1-\gamma  \Big) \Big\} \nonumber\\
&&\times L_{\psi _1}\Big( w_2= z(1-p^2)\Big\{ (1-t)+ \frac{tsv}{(1-v)}\Big\} \Big) \label{eq:168b}
\end{eqnarray}
\begin{eqnarray}
\sideset{_{}^{}}{_{\psi _0}^{\psi _1}}\bigwedge \big(\gamma ;z\big) &=& \frac{2\;\psi _0!}{(2\pi i)^2 B(\frac{1}{2},\psi _1+1)}\int_0^{\infty } ds \;s^{-\frac{1}{2}}(1+s)^{-(\psi _1+\frac{3}{2})}\int_0^1 dt\; t^{\frac{1}{2}-\gamma }\nonumber\\
&&\times \oint \frac{dv}{v^{\psi _0+1}(1-v)^{2-\gamma }} \oint \frac{du}{u^{\psi _1+1}(1-u)} \Bigg\{ \frac{zstvu}{(1-v)(1-u)} \label{eq:168c}\\
&&\times \Bigg( \frac{1}{3}M\Big(1,\frac{5}{2},w_3\Big)-M\Big(1,\frac{3}{2},w_3\Big) \Bigg)+ \Big( \frac{\omega }{2}+1-\gamma  \Big) M\Big(1,\frac{3}{2},w_3\Big)\Bigg\} \nonumber\\
&&\;  \mbox{where}, w_3=\frac{zu}{(1-u)}\Big\{ -(1-t)-\frac{stv}{(1-v)}\Big\}\nonumber
\end{eqnarray}
\begin{eqnarray}
\sideset{_{}^{}}{_{\psi _0}^{\psi _1}}\bigwedge \big(\gamma ;z\big) &=& \frac{\psi _0!}{(2\pi i)^2 B(\frac{1}{2},\psi _1+1)}\int_0^{\infty } ds \;s^{-\frac{1}{2}}(1+s)^{-(\psi _1+\frac{3}{2})}\int_0^1 dp\;(1-p^2)^{-\frac{1}{2}} \nonumber\\
&&\times  \oint \frac{dv}{v^{\psi _0+1}(1-v)^{2-\gamma }} \oint \frac{du e^{-\frac{zup}{(1-u)}}}{u^{\psi _1+1}(1-u)} \Bigg\{ -\frac{zspvu}{(1-v)(1-u)} \label{eq:168d}\\
&&\times \frac{1}{(\frac{5}{2}-\gamma )}M\Big(\frac{5}{2}-\gamma ,\frac{7}{2}-\gamma ,w_4\Big) + \frac{\Big( \frac{\omega }{2}+1-\gamma  \Big)}{(\frac{3}{2}-\gamma )}M\Big(\frac{3}{2}-\gamma ,\frac{5}{2}-\gamma ,w_4\Big)\Bigg\} \nonumber\\ 
&&\; \mbox{where}, w_4=\frac{zup}{(1-u)}\Big\{1-\frac{sv}{(1-v)}\Big\} \nonumber
\end{eqnarray}
\begin{eqnarray}
\sideset{_{}^{}}{_{\psi _0}^{\psi _1}}\bigwedge \big(\gamma ;z\big) &=& \frac{\psi _0!\;\sqrt{\pi }}{(2\pi i)^2 B(\frac{1}{2},\psi _1+1)}\int_0^{1 } dt \;t^{\frac{1}{2}-\gamma } \int_{-1}^1 dp \oint \frac{dv}{v^{\psi _0+1}(1-v)^{2-\gamma }}\nonumber\\
&& \times  \oint du\frac{ e^{-\frac{zu}{(1-u)}(1-t)(1-p^2)}}{u^{\psi _1+1}(1-u)} \Bigg\{ -\frac{ztvu(1-p^2)}{2(1-v)(1-u)} U\Big(\frac{3}{2},-\psi _1+1,w_5\Big)\nonumber\\
&&+ \Big( \frac{\omega }{2}+1-\gamma  \Big) U\Big(\frac{1}{2},-\psi _1,w_5\Big)\Bigg\}\nonumber\\
&& \mbox{where}, w_5=\frac{ztuv}{(1-v)(1-u)}(1-p^2)
\label{eq:168e}
\end{eqnarray}
And its integral representation is
\begin{eqnarray}
 \mathcal{RW}_{\psi _0,\psi _1}\Big( \gamma =\frac{1}{2}(1+\nu ); z=-\frac{1}{2}\mu x^2\Big) &=& z^{1-\gamma }\Bigg\{ \frac{\psi _0!}{2\pi i}  \oint dv \frac{e^{-\frac{zv}{(1-v)}}}{v^{\psi _0+1}(1-v)^{2-\gamma }}\nonumber\\
&&- \frac{\varepsilon }{2} x  \sideset{_{}^{}}{_{\psi _0}^{\psi _1}}\bigwedge \big(\gamma ;z\big)\Bigg\}
 \label{eq:169}
\end{eqnarray}
where,
\begin{eqnarray}
 \sideset{_{}^{}}{_{\psi _0}^{\psi _1}}\bigwedge \big(\gamma ;z\big) &=& \frac{\psi _0!}{(2\pi i)^2 B(\frac{1}{2},\psi _1+1)}\int_0^{\infty } ds \;s^{-\frac{1}{2}}(1+s)^{-(\psi _1+\frac{3}{2})}\int_0^1 dt\; t^{\frac{1}{2}-\gamma } \int_{-1}^1 dp \nonumber\\
&&\times \oint \frac{dv}{v^{\psi _0+1}(1-v)^{2-\gamma }}\oint \frac{du}{u^{\psi _1+1}(1-u)} \bigg\{-\frac{zstuv(1-p^2)}{(1-u)(1-v)}\nonumber\\
&&+\Big( \frac{\omega }{2}+1-\gamma  \Big) \bigg\} e^{-\frac{zu}{(1-u)}(1-p^2)\big\{ (1-t)+ \frac{tsv}{1-v}\big\}}
\label{eq:170}
\end{eqnarray}
 Due to space restriction proofs of (\ref{eq:166})-(\ref{eq:170}) are not included in the paper, but feel free to contact me for the proofs. 
\subsubsection{Generating function}
Let's try to construct the generating function of the $2^{\mbox{nd}}$ kind of GCH function. First, multiply ${\displaystyle \sum_{\psi _0=0}^{\infty } \sum_{\psi _1=\psi _0}^{\infty } \frac{B\big(\psi _1+1, \frac{1}{2}\big)}{\psi _0!} \;v_0^{\psi _1} v_1^{\psi_0 }}$ on both sides of integral form of $2^{\mbox{nd}}$ kind of GCH function in (\ref{eq:166})-(\ref{eq:300}) where $|v_0|<1$ and $|v_1|<1$. Then, its solution is the following way.
\begin{eqnarray}
&& \sum_{\psi _0=0}^{\infty } \sum_{\psi _1=\psi _0}^{\infty } \frac{B\big(\psi _1+1, \frac{1}{2}\big)}{\psi _0!} \;v_0^{\psi _1} v_1^{\psi_0 }\; \mathcal{RW}_{\psi _0,\psi _1}\Big(\gamma =\frac{1}{2}(1+\nu ); z=-\frac{1}{2}\mu x^2\Big)\nonumber\\ 
&=& z^{1-\gamma } \Bigg\{ \sum_{m=0}^{\infty } \frac{\Gamma (\frac{1}{2}) (-zv_0v_1)^m}{\Gamma (m+\frac{3}{2})} F_1 \big(m+1 ;1 ,2-\gamma +m ;m+\frac{3}{2} ;v_0,v_0 v_1 \big) \nonumber\\
&&-\frac{\varepsilon }{2} x \bigg\{ \sum_{n,m=0}^{\infty } \frac{\Gamma (\frac{1}{2}) \Gamma (\frac{3}{2}) (n+m+1)(n+1)_m (1-v_1)^{-(n+3-\gamma )}(-zv_0)^{n+m+1}(v_1)^{n+1}}{(n+\frac{5}{2}-\gamma )\Gamma (m+\frac{5}{2}) \Gamma (n+m+\frac{5}{2}) }\nonumber\\
&&\times   F_1 \Big(m+1; n+m+2, -n;m+\frac{5}{2}; v_0,\frac{1}{v_1}\Big)\nonumber\\
&&+ \Big(\frac{\omega }{2}+1-\gamma \Big) \sum_{n,m=0}^{\infty } \frac{\big(\Gamma (\frac{1}{2})\big)^2 (n+1)_m (1-v_1)^{-(n+2-\gamma)}(-zv_0)^{n+m}(v_1)^{n}}{(n+\frac{3}{2}-\gamma ) \Gamma (m+\frac{3}{2})\Gamma (n+m+\frac{3}{2}) } \nonumber\\ 
&&\times  F_1 \Big(m+1; n+m+1, -n;m+\frac{3}{2}; v_0,\frac{1}{v_1}\Big)\bigg\} \Bigg\} 
\label{eq:171}
\end{eqnarray}
And its integral representation of generating function is
\begin{eqnarray}
&& \sum_{\psi _0=0}^{\infty } \sum_{\psi _1=\psi _0}^{\infty } \frac{B\big(\psi _1+1, \frac{1}{2}\big)}{\psi _0!} \;v_0^{\psi _1} v_1^{\psi_0 }\; \mathcal{RW}_{\psi _0,\psi _1}\Big(\gamma =\frac{1}{2}(1+\nu ); z=-\frac{1}{2}\mu x^2\Big)\nonumber\\
&=&  z^{1-\gamma } \Bigg\{ \int_0^1 dt\; (1-t)^{-\frac{1}{2}}(1-v_0t)^{-1} (1-v_0v_1 t)^{\gamma-2 } e^{-\frac{zv_0v_1t}{1-v_0v_1t}}\nonumber\\
&&-\frac{\varepsilon }{2} \frac{x}{(1-v_1)^{2-\gamma }}\int_0^{1} dy \;(1-y)^{-\frac{1}{2}}(1-v_0 y)^{-1}\label{eq:172}\\
&&\times \int_0^1 dt\; t^{\frac{1}{2}-\gamma } \int_{-1}^1 dp\;\bigg\{ \Big( \frac{\omega }{2}+1-\gamma \Big) - \frac{zv_0 v_1t(1-y)}{(1-v_1)(1- v_1 y)}\bigg\} e^{-\frac{z v_0 y}{1- v_0 y}\Big\{1-\frac{t(y-v_1)}{y(1-v_1)}\Big\}(1-p^2)} \Bigg\}\nonumber
\end{eqnarray}
Due to space restriction proofs of (\ref{eq:171}) and (\ref{eq:172}) are not included in the paper, but feel free to contact me for the proofs. 
\subsubsection{Orthogonal relation} 
By using similar process as I construct the orthogonal relation of the first independent solution of GCH function, the orthogonal relation of the second independent solution of GCH function is
\begin{eqnarray}
&& \int_0^{\infty }dx\;x^{\nu }e^{\frac{1}{2}\mu x^2+\varepsilon x} \mathcal{RW}_{\psi _0,\psi _1}(z) \mathcal{RW}_{\phi _0,\phi _1}(z) \nonumber\\ 
&\approx & \Bigg\{ \frac{|\mu |^{4-3\gamma}}{2^{1-\gamma}} \frac{\Gamma (2-\gamma )B(3-\gamma ,\psi _0)}{B(6-3\gamma ,\psi _0)}
+ \varepsilon \frac{(-1)^{\psi _0}2^{\gamma -\frac{1}{2}} }{|\mu |^{\gamma +\frac{1}{2}}} \Bigg\{\frac{\Gamma \big(\psi _0+ \frac{3}{2}-\gamma  \big)\Gamma \big(\psi _0+ \frac{5}{2}-\gamma  \big)}{\Gamma \big( \frac{3}{2}-\gamma \big)} \nonumber\\ 
&&-  \frac{\Gamma (\psi _0+2-\gamma )}{\Gamma (2-\gamma )}\sum_{n=0}^{\psi _0} \frac{(-\psi _0)_n}{ n!(2-\gamma )_n} \label{eq:178}\\ 
&&\times \sum_{k=0}^{\psi _1 -n} \frac{(\frac{\omega }{2}+n+1-\gamma ) \Gamma (n+\frac{1}{2})\Gamma (n+\frac{3}{2}-\gamma)(n-\psi _1)_k}{\Gamma (n+k+\frac{3}{2}-\psi _0)} \Bigg\}\Bigg\} \delta _{\psi _0,\phi _0} \;\delta _{\psi _1,\phi _1} \nonumber
\end{eqnarray}
If there is an analytic function $\Phi (x)$ having first and second continuous derivatives in $[0,\infty ]$ and approaching zero when $x\rightarrow \infty $, it can be expanded in terms of $\mathcal{RW}_{\psi _0,\psi _1}( \gamma ; z)$.
\begin{equation}
\Phi(x) = \sum_{\psi _0 =0}^{\infty } \sum_{\psi _1=\psi _0}^{\infty } B_{\psi _0,\psi _1} \mathcal{RW}_{\psi _0,\psi _1}\Big( \gamma =\frac{1}{2}(1+\nu ); z=-\frac{1}{2}\mu x^2\Big)\label{eq:179}
\end{equation}
where,
\begin{eqnarray}
B_{\psi _0,\psi _1} &\approx & \int_{0}^{\infty } dx\;x^{\nu } e^{\frac{1}{2}\mu x^2+ \varepsilon x}\mathcal{RW}_{\psi _0,\psi _1}( \gamma ; z) \Phi (x) \Bigg[  \frac{|\mu |^{4-3\gamma}}{2^{1-\gamma}} \frac{\Gamma (2-\gamma )B(3-\gamma ,\psi _0)}{B(6-3\gamma ,\psi _0)}\nonumber\\
&&+ \varepsilon \frac{(-1)^{\psi _0}2^{\gamma -\frac{1}{2}} }{|\mu |^{\gamma +\frac{1}{2}}} \Bigg\{\frac{\Gamma \big(\psi _0+ \frac{3}{2}-\gamma  \big)\Gamma \big(\psi _0+ \frac{5}{2}-\gamma  \big)}{\Gamma \big( \frac{3}{2}-\gamma \big)} - \frac{\Gamma (\psi _0+2-\gamma )}{\Gamma (2-\gamma )}\sum_{n=0}^{\psi _0} \frac{(-\psi _0)_n}{ n!(2-\gamma )_n} \nonumber\\
&&\times \sum_{k=0}^{\psi _1 -n} \frac{(\frac{\omega }{2}+n+1-\gamma ) \Gamma (n+\frac{1}{2})\Gamma (n+\frac{3}{2}-\gamma)(n-\psi _1)_k}{\Gamma (n+k+\frac{3}{2}-\psi _0)} \Bigg\}  \Bigg]^{-1} \label{eq:180}
\end{eqnarray}
Due to space restriction proofs of (\ref{eq:178})-(\ref{eq:180}) are not included in the paper, but feel free to contact me for the proofs. 
\section{\label{sec:level4}Infinite series for $\nu $= non-integer}
In the previous section, I discussed the solutions of polynomial case. 

The $1^{st}$ independent solution has two eigenvalues which are
\begin{equation}
\begin{cases} 
|\alpha _0|= -\frac{\Omega }{2\mu } \Rightarrow 0,1,2,3,\cdots \cr
|\alpha _1|= -\frac{\Omega }{2\mu }-\frac{1}{2} \Rightarrow 0,1,2,3,\cdots \hspace{1cm}; \mbox{only}\;\mbox{if}\; |\alpha _1|\geq |\alpha _0|
\end{cases}
\label{eq:211}
\end{equation}
Also, the $2^{nd}$ independent solution has two eigenvalues which are
\begin{equation}
\begin{cases} 
\psi _0= -\frac{\Omega }{2\mu }+\gamma -1 \Rightarrow 0,1,2,3,\cdots \cr
\psi _1= -\frac{\Omega }{2\mu }+\gamma -\frac{3}{2} \Rightarrow 0,1,2,3,\cdots \hspace{1cm}; \mbox{only}\;\mbox{if}\; \psi _1\geq \psi _0
\end{cases}
\label{eq:212}
\end{equation}
These conditions make the solutions as polynomial. 

However, if
\begin{equation}
\begin{array}{ll}
-\frac{\Omega }{2\mu } \ne  \left\{\begin{array}{c}
0,1,2,3, \cdots \\
\frac{1}{2}, \frac{3}{2}, \frac{5}{2}, \cdots
\end{array} \right.
\hspace{2cm}-\frac{\Omega }{2\mu } \ne 
\left\{\begin{array}{c}
1-\gamma ,2-\gamma ,3-\gamma , \cdots \\
\frac{3}{2}-\gamma ,\frac{5}{2}-\gamma ,\frac{7}{2}-\gamma , \cdots
\end{array}\right.
\end{array}
\label{eq:213}
\end{equation}
Then analytic solutions of GCH function turn to be infinite series. 
 
From the $1^{st}$ kind of GCH polynomial, let $|\alpha _0|= -\frac{\Omega }{2\mu }$ and $|\alpha _1|= -\frac{\Omega }{2\mu }-\frac{1}{2}$ in (\ref{eq:82}), (\ref{eq:83a}) and (\ref{eq:83b}) where $|\alpha _0|,|\alpha _1|\ne 0,1,2,3,\cdots$. The series turns to the infinite series for the case of $\lambda _1 = 0$. I denote the $1^{st}$ independent solution of GCH function for the infinite series as $\mathcal{QW}\Big( \gamma =\frac{1}{2}(1+\nu ); z=-\frac{1}{2}\mu x^2\Big)$. 

From the $2^{nd}$ kind of GCH polynomial, let $\psi _0 = -\frac{\Omega }{2\mu }-1+\gamma $ and $\psi _1 = -\frac{\Omega }{2\mu }-\frac{1}{2}+\gamma $ in (\ref{eq:166})-(\ref{eq:300}) where $\psi _0,\psi _1 \ne 0,1,2,3,\cdots$. The series turns to be the infinite series for the case of $\lambda _2 = 1-\nu$. I denote the $2^{nd}$ independent solution of GCH function for the infinite series as $\mathcal{RW}\Big( \gamma =\frac{1}{2}(1+\nu ); z=-\frac{1}{2}\mu x^2\Big)$. Due to space restriction power series expansions in closed forms of GCH function for the infinite series are not included in the paper, but feel free to contact me for more details. 

 When $\nu $ is integer, one of two solution of the GCH equation does not have any meaning, because $\mathcal{RW}_{\psi _0,\psi _1}\Big( \gamma =\frac{1}{2}(1+\nu ); z=-\frac{1}{2}\mu x^2\Big)$ and $\mathcal{RW}\Big( \gamma =\frac{1}{2}(1+\nu ); z=-\frac{1}{2}\mu x^2\Big)$ can be described as $\mathcal{QW}_{|\alpha _{0}|,|\alpha _{1}|}\Big( \gamma =\frac{1}{2}(1+\nu ); z=-\frac{1}{2}\mu x^2\Big)$ and $\mathcal{QW}\Big( \gamma =\frac{1}{2}(1+\nu ); z=-\frac{1}{2}\mu x^2\Big)$ as long as $|\lambda _1-\lambda _2|=|\nu -1|=$ integer. It is required that $\gamma \ne 0,-1,-2,\cdots$ for the first kind of independent solution of GCH function for the polynomial and infinite series. Also $\gamma \ne 2,3,4,\cdots$ is required for the second kind of independent solution of GCH function for both the polynomial and infinite series.  

Due to space restriction I do not include a polynomial and an infinite series of GCH equation for $\nu \in \mathbb{Z}$ in this paper. If the time is permitted, I will publish the power series expansion in closed forms and its integral forms of GCH function for $\nu \in \mathbb{Z}$.
\section{Conclusion}
Many great scholars noticed that the excited meson including baryon conduct like a elongated bilocal linear structure hold by the gluon flux tube behaving as a scalar linear potential for high rotational excitation at large separation.
G\"{u}rsey and collaborators wrote the semi-relativistic Hamiltonian for the $q-\bar{q}$ system neglecting small mass of quarks\cite{1991}, suggested by Lichitenberg \textit{et al.}\cite{Lich1982}. In this system, QCD forces are flavor independent, the strong-coupling potential like a Coulomb potential is negligible and the confining part of QCD potential is spin independent.    
They detected that their wave equation neglecting the mass of quark is equivalent confluent hypergeometric differential equation in which the recursive relation involves two terms in its power series expansion. They only obtained one mass formula for meson with the universal Regge slope. Their slope is quiet matched with experimental results for the high rotationally excited meson and baryon.

However, since I include the mass of quark into their spin free Hamiltonian involving only scalar potential, the new type ordinary differential equation arises, designated as grand confluent hypergeometric (GCH) equation. Its recurrence relation consists of three terms in the formal series solution.  
By using Frobenius method and putting the power series expansion into the GCH ordinary differential equation, three recursive relation of coefficients starts to appear. Currently the analytic solution of three term recurrence relation is unknown. Due to its complexity for mathematical computations, more than three term case has been neglected. In this paper I claim that $\varepsilon$, one of coefficients, is extremely small quantity because of its physical meaning: $\varepsilon$ is correspondent to the small mass of quark. Therefore I construct the analytic solution of GCH function including only up to the first order of $\varepsilon$. More than second order of $\varepsilon$ is negligible.  

From the above all, I show asymptotic expansions of GCH function for an infinite series and a polynomial. I construct the power series expansions in closed forms of the GCH polynomial including only up to the first order of $\varepsilon$. For a formal series solution of the GCH equation for an infinite series, it is available as arXiv.\cite{Chou2012a} As we see all solutions of power series expansions in the GCH polynomial, denominators and numerators in $g(x)_{domin.}$ and $g(x)_{small}$ terms arise with Pochhammer symbol: the meaning of this is that the analytic solutions of the GCH ordinary differential equation can be described as Hypergoemetric-type function in a strict mathematical way. 

I construct representations in closed form integrals of the GCH polynomial in an easy way since I have power series expansions with Pochhammer symbols in numerators and denominators.
I show that the first kind of Confluent hypergeometric polynomial appears in the sub-integral forms of the GCH polynomial in (\ref{eq:108}). 
The Confluent hypergeometric function in the sub-integral forms of the GCH function is able to be transformed to other well-known special functions analytically such as the first and second Kummer and Laguerre functions.\cite{Chou2012a} Understanding the connection between other special functions is important in the mathematical and physical points of views as we all know.

Analytic integral forms including only up to the first order of $\varepsilon$ of GCH functions are derived from power series expansion in closed forms of the GCH differential equation. I construct the generating functions of the GCH polynomial from the its integral forms and I derive orthogonal relations of GCH polynomial including only up to the first order of $\varepsilon$. 
The orthogonal relation of the GCH polynomial is important in the physical point of view because we can derive recurrence relations and expectation values of physical quantities from its orthogonal relation. For the case of hydrogen-like atoms, the normalized wave function is derived from the generating function of associated Laguerre polynomial. The expectation value of physical quantities such as position and momentum is constructed by applying the recursive relation of associated Laguerre polynomial with its generating function. 

The most remarkable result in this paper is that infinite eigenvalues (the mass formula for the meson for high rotational excitation) is constructed analytically in (\ref{eq:35a})-(\ref{eq:35e}) since the mass of quark is included in a semi-relativistic wave equation involving only scalar potential with spin independent for the wave function. However, G\"{u}rsey and his collaborators obtained only one eigenvalue neglecting the mass of quark due to its complicated mathematical calculations. 
In the mathematical points of view, infinite eigenvalues arise because of the three term recursion relation in a power series expansion for a polynomial in their wave equation: for example, Heun, confluent Heun, the GCH (later on, I noticed that the GCH equation is the more general form of Biconfluent Heun equation\cite{Chou2012i,Chou2012j}), Lam\'{e}, Mathieu equations, etc are composed of three term recurrence relation in their Frobenius solutions. In contrast, two term recurrence relation in a formal series solution for a polynomial only has one eigenvalue such as hypergeometric, confluent hypergeometric, Laguerre, Legendre differential equations, etc.

 Because of the above mathematical phenomenon, many linearly increasing lines with same slopes and intercepts including bunch of eigenvalues arise. As you see the QCD experiments for the high rotationally excited meson and baryon, there are only finite linearly increasing lines. Theoretically, infinite linearly increasing lines can be created in hadron experiments. The reason why we have only finite increasing straight lines is that the full design power per beam of the Large Hadron Collider (LHC) is not strong enough. If we input extreme huge power in the LHC, we will probably obtain uncountable straight lines for the excited meson. And its results might be possible to be matched with (\ref{eq:35a})-(\ref{eq:35e}). In future papers, I will show mass formulas for $q-qq$ baryon, $\bar{q}-\bar{q}\bar{q}$ antibaryon and  $qq-\bar{q}\bar{q}$ exotic meson (Tetraquark) including their semi-relativistic normalized wave functions neglecting the fourth component of a vector potential with high rotational excitations for a large separation.  

Furthermore, I generalize three term recurrence relation in linear differential equations. I derive analytic solutions having three different successive coefficients in its formal series for polynomials and infinite series \cite{Chou2012b}. I show how to solve mathematical equations having three term recursion relation and go on producing analytic solutions of some of well known special function theories that include Heun \cite{Chou2012c,Chou2012d}, Mathieu\cite{Chou2012e}, Lam\'{e}\cite{Chou2012f,Chou2012g,Chou2012h} and GCH Functions\cite{Chou2012i,Chou2012j}. I hope these new functions and their solutions will produce remarkable new range of applications not only in supersymmetric field theories as is shown here, but in the areas of all different classes of mathematical physics, applied mathematics and in engineering applications.  
\vspace{5mm}

\section{Series ``Special functions and three term recurrence formula (3TRF)''} 

This paper is 1st out of 10.
\vspace{3mm}

1. ``Approximative solution of the spin free Hamiltonian involving only scalar potential for the $q-\bar{q}$ system'' \cite{Chou2012a} - In order to solve the spin-free Hamiltonian with light quark masses we are led to develop a totally new kind of special function theory in mathematics that generalize all existing theories of confluent hypergeometric types. We call it the Grand Confluent Hypergeometric Function. our new solution produces previously unknown extra hidden quantum numbers relevant for description of supersymmetry and for generating new mass formulas.
\vspace{3mm}

2. ``Generalization of the three-term recurrence formula and its applications'' \cite{Chou2012b} - Generalize three term recurrence formula in linear differential equation.  Obtain the exact solution of the three term recurrence for polynomials and infinite series.
\vspace{3mm}

3. ``The analytic solution for the power series expansion of Heun function'' \cite{Chou2012c} -  Apply three term recurrence formula to the power series expansion in closed forms of Heun function (infinite series and polynomials) including all higher terms of $A_n$'s.
\vspace{3mm}

4. ``Asymptotic behavior of Heun function and its integral formalism'', \cite{Chou2012d} - Apply three term recurrence formula, derive the integral formalism, and analyze the asymptotic behavior of Heun function (including all higher terms of $A_n$'s). 
\vspace{3mm}

5. ``The power series expansion of Mathieu function and its integral formalism'', \cite{Chou2012e} - Apply three term recurrence formula, analyze the power series expansion of Mathieu function and its integral forms.  
\vspace{3mm}

6. ``Lam\'{e} equation in the algebraic form'' \cite{Chou2012f} - Applying three term recurrence formula, analyze the power series expansion of Lam\'{e} function in the algebraic form and its integral forms.
\vspace{3mm}

7. ``Power series and integral forms of Lam\'{e} equation in   Weierstrass's form and its asymptotic behaviors'' \cite{Chou2012g} - Applying three term recurrence formula, derive the power series expansion of Lam\'{e} function in   Weierstrass's form and its integral forms. 
\vspace{3mm}

8. ``The generating functions of Lam\'{e} equation in   Weierstrass's form'' \cite{Chou2012h} - Derive the generating functions of Lam\'{e} function in   Weierstrass's form (including all higher terms of $A_n$'s).  Apply integral forms of Lam\'{e} functions in   Weierstrass's form.
\vspace{3mm}

9. ``Analytic solution for Grand Confluent Hypergeometric function'' \cite{Chou2012i} - Apply three term recurrence formula, and formulate the exact analytic solution of Grand Confluent Hypergeometric function (including all higher terms of $A_n$'s). Replacing $\mu $ and $\varepsilon \omega $ by 1 and $-q$, transforms the grand confluent hypergeometric function into Biconfluent Heun function.
\vspace{3mm}

10. ``The integral formalism and the generating function of Grand Confluent Hypergeometric function'' \cite{Chou2012j} - Apply three term recurrence formula, and construct an integral formalism and a generating function of Grand Confluent Hypergeometric function (including all higher terms of $A_n$'s). 
\section*{Acknowledgment}
I thank Bogdan Nicolescu. The discussions I had with him on number theory was of great joy.

\end{document}